\documentclass[preprint,nofootinbib]{revtex4}

\usepackage{amsmath}
\usepackage{graphicx}

\renewcommand{\Im}{\mathcal{I}m}

\newcommand{\beq}{\begin{equation}}
\newcommand{\eeq}{\end{equation}}
\newcommand{\beqa}{\begin{eqnarray}}
\newcommand{\eeqa}{\end{eqnarray}}
\newcommand{\no}{\nonumber}
\def\OMIT#1{{}}
\newcommand{\lsim}{\mathrel{\rlap{\lower4pt\hbox{\hskip1pt$\sim$}}
    \raise1pt\hbox{$<$}}}         %less than or approx. symbol
\newcommand{\gsim}{\mathrel{\rlap{\lower4pt\hbox{\hskip1pt$\sim$}}
    \raise1pt\hbox{$>$}}}         %greater than or approx. symbol

\begin{document}

%\preprint{{\vbox{\hbox{}\hbox{}\hbox{}
%    \hbox{WIS/10/05-May-DPP}
%\hbox{hep-ph/0609178}}}}

\vspace*{.0cm}

\title{New Physics and CP Violation\\ in Singly Cabibbo Suppressed $D$
Decays}

\author{Yuval Grossman}\email{yuvalg@physics.technion.ac.il}
\affiliation{Department of Physics, Technion-Israel
  Institute of Technology, Technion City, Haifa 32000,
  Israel}
\author{Alexander L. Kagan}\email{kagan@physics.uc.edu}
\affiliation{Department of Physics,
  University of Cincinnati, Cincinnati, Ohio 45221, U.S.A.}
\author{Yosef Nir}\email{yosef.nir@weizmann.ac.il}
\affiliation{Department of Particle Physics,
  Weizmann Institute of Science, Rehovot 76100, Israel\vspace*{3mm}}

%\date{\today}
%\pacs{}

\begin{abstract}\vspace*{3mm}
We analyze various theoretical aspects of CP violation in singly
Cabibbo suppressed (SCS) $D$-meson decays, such as $D \to K K,\pi
\pi$.  In particular, we explore the possibility that CP asymmetries
will be measured close to the present level of experimental
sensitivity of ${\cal O}(10^{-2})$. Such measurements would signal new
physics. We make the following points: (i) The mechanism at work in
neutral $D$ decays could be indirect or direct CP violation (or
both). (ii) One can experimentally distinguish between these
possibilities. (iii) If the dominant CP violation is indirect, then
there are clear predictions for other modes.  (iv) Tree-level direct
CP violation in various known models is constrained to be much smaller
than $10^{-2}$.  (v) SCS decays, unlike Cabibbo favored or doubly
Cabibbo suppressed decays, are sensitive to new contributions from QCD
penguin operators and especially from chromomagnetic dipole
operators. This point is illustrated with supersymmetric gluino-squark
loops, which can yield direct CP violating effects of ${\cal
O}(10^{-2})$.
\end{abstract}

\maketitle

%%%%%%%%%%%%%%%%%%%%%%%%%%%%
\section{Introduction}
\label{sec:introduction}
CP violation in $D$-meson decays provides a unique probe of new
physics. First, the Standard Model predicts very small effects,
smaller than ${\cal O}(10^{-3})$, so that a signal at the present
level of experimental sensitivity
\cite{Aubert:2003pz,Abe:2003ys,Aubert:2005gj,Csorna:2001ww,Acosta:2004ts,Bonvicini:2000qm,Link:2001zj},
${\cal O}(10^{-2})$, would clearly signal new physics. Second, the
neutral $D$ system is the only one where the external up-sector quarks
are involved.  Thus it probes models in which the up sector plays a
special role, such as supersymmetric models with alignment
\cite{Nir:1993mx,Leurer:1993gy} and, more generally, models in which
CKM mixing is generated in the up sector. Third, singly Cabibbo
suppressed (SCS) decays are sensitive to new physics contributions to
penguin and dipole operators.

Let us elaborate on the first point, that is, the smallness of CP
violation within the Standard Model (SM). The basic argument is that
the physics of both $D^0-\overline{D}^0$ mixing and SCS $D$ decays
involves, to an excellent approximation, only the first two quark
generations and is therefore CP conserving \cite{Bianco:2003vb}. In
other words, SM CP violation in these decays is CKM suppressed. As
concerns the $D^0-\overline{D}^0$ mixing amplitude, SM CP violation
enters at ${\cal O}[|(V_{cb}V_{ub})/(V_{cs}V_{us})|]
\sim10^{-3}$. Furthermore, this suppression is relative to the short
distance contribution, which is known to lie well below the present
experimental sensitivity. (The SM contribution could saturate the
present bounds on $y$ \cite{Falk:2001hx} and $x$ \cite{Falk:2004wg},
but this would necessarily be due to the long distance contribution.)
The CP violation contribution to the $c\to u\bar ss$ and $c\to u\bar
dd$ decays is both CKM- and loop-suppressed and, therefore, entirely
negligible. We conclude that CP violation in SCS $D$ decays at the
percent level signals new physics
\cite{Bigi:1986dp,Blaylock:1995ay,Bergmann:2000id}.

As concerns the third point, among all hadronic $D$ decays, the    
SCS decays are uniquely sensitive to CP violation in $c\to u
\bar q q $ transitions and, consequently, to new contributions to
the $\Delta C=1$ QCD penguin and chromomagnetic dipole operators. In
particular, such contributions can affect neither the Cabibbo favored 
($c\to s \bar d u$) nor the doubly Cabibbo suppressed ($c \to d \bar s
u$) decays.

In Sections \ref{sec:indirect} and \ref{sec:directvsin} we present the
formalism of CP violation in SCS $D$ decays. For final CP eigenstates,
indirect CP violation is universal.  Thus, for example, equal
time-integrated CP asymmetries in $D \to K^+ K^- $ and $D \to \pi^+
\pi^- $ would be a signal for indirect CP violation.  By combining
time-dependent and time-integrated measurements it is possible to
separate out the universal indirect and generally non-universal direct
CP asymmetry contributions. In the case of final non-CP eigenstates,
such as $\rho^\pm \pi^\mp $ or $K^{*\pm } K^{\mp}$, a Dalitz plot
analysis allows one to further separate out the indirect CP
asymmetries originating from CP violation in mixing and from CP
violation in the interference of decays with and without mixing, and
to separately determine the neutral $D$-meson mass and lifetime
differences, up to discrete ambiguities.

In Sections \ref{sec:directtree} and \ref{sec:directloop} we discuss
direct CP violation. In Section \ref{sec:directtree}, we survey
models which give rise to direct CP violation in SCS decays via
{\it tree}-level decay amplitudes, e.g., flavor-changing
$Z$ or $Z^\prime$ couplings or supersymmetric R-parity violating
couplings. We find that typically these contributions are constrained
to lie well below the present experimental sensitivity.

In Section \ref{sec:directloop} we discuss {\it loop}-induced
effects. Here the situation is different, as direct CP violation at
the level of $10^{-2}$ is often allowed and, for specific models, even
expected. Two specific supersymmetric examples employing
up-squark/gluino loops are discussed: contributions to the dipole
operators due to flavor-changing ``left-right'' (LR) squark mixing,
and contributions to the QCD penguin and dipole operators due to
flavor-changing ``left-left'' (LL) squark mixing. Remarkably, we find
that LR squark mixing can yield direct CP violation at the current
level of sensitivity, while indirect CP violation remains
negligible. The key factor is a strong enhancement of the requisite
quark chirality flip in the dipole operators by a factor $m_{\tilde g}
/m_c $ which is absent in the mixing amplitude. For LL squark mixing,
annihilation leads to an order of magnitude uncertainty in the QCD
penguin operator matrix elements, so that direct CP asymmetries of
${\cal O}(10^{-2})$ cannot be ruled out. In this case, however, 
indirect CP violation is also non-negligible. Implications for CP violation
in supersymmetric flavor models with alignment, which predict the orders of
magnitude of the LR and LL squark mixings, are discussed.

In this analysis, some hadronic subtleties are involved. We employ
naive factorization to evaluate the impact of new contributions to the
QCD penguin operators, and QCD factorization \cite{Beneke:2001ev} to
estimate the contributions of chromomagnetic dipole operators. We
argue that there is a large theoretical uncertainty related to
annihilation in both (SM) tree and (new physics) penguin
contributions: Experimental information as well as hadronic models
lead us to think that annihilation could play a prominent role and, in
particular, strongly enhance the latter.  Details are provided in the
Appendix.  Finally, isospin invariance and, to a lesser extent,
$U$-spin invariance of the gluonic transitions predict patterns of
direct CP violation among various SCS decay modes.  These can be used
to test for new contributions to the QCD penguin and dipole operators.

We conclude in Section \ref{sec:conclusions} with a summary of our
results and a brief discussion of additional decay modes which will be
useful for learning about the possible intervention of new physics in
SCS $D$ meson decays.

%%%%%%%%%%%%%%%%%%%%
\section{Formalism}
\label{sec:indirect}
The SCS decays, $c\to u\bar ss$ and $c\to u\bar dd$, lead to final
states that are common to $D^0$ and $\overline{D}^0$. These could be
CP eigenstates (such as $K^+K^-$, $\pi^+\pi^-$, $\phi\pi^0$ and
$\rho^0\phi^0$), or non-CP eigenstates (such as $\rho^+\pi^-$,
$K^{*+}K^-$ and $K^{*0}K_S$). 

We use the following standard notations:
\beqa
\tau&\equiv&\Gamma_D t,\qquad
\Gamma_D\equiv\frac{\Gamma_{D_H}+\Gamma_{D_L}}{2}, \no\\
A_f&\equiv&A(D^0\to f),\qquad
 \overline{A}_f\equiv
A(\overline{D}{}^0\to f),\no\\
A_{\bar f}&\equiv&A(D^0\to \overline f),\qquad
 \overline{A}_{\overline f}\equiv A(\overline{D}{}^0\to \overline f),\no\\
x&\equiv&\frac{\Delta
  m_D}{\Gamma_D}\equiv\frac{m_{D_H}-m_{D_L}}{\Gamma_D},\qquad
y\equiv\frac{\Delta\Gamma_D}{2\Gamma_D}\equiv
\frac{\Gamma_{D_H}-\Gamma_{D_L}}{2\Gamma_D},\no\\
\lambda_f&\equiv&\frac qp \frac{\overline{A}_f}{A_f}, \qquad
R_m \equiv\left|\frac qp\right|,\qquad
R_f\equiv\left|{\overline{A}_f \over A_f}\right|.
\eeqa
Here $D_H$ and $D_L$ stand for the heavy and light mass eigenstates, 
and $q$ and $p$ are defined via $|D_{H,L}\rangle=p|D^0\rangle\mp
q|\overline{D}^0\rangle$.

The time dependent decay rates into a final state $f$ can be
written as follows (see, for example, \cite{Nir:2005js}): 
\beqa\label{tddr}
\Gamma(D^0(t)\to f)&=&e^{-\tau}|A_f|^2
\Big\{(1+|\lambda_f|^2) \cosh(y\tau)
+(1-|\lambda_f|^2) \cos(x\tau)\nonumber \\
&&+2 {\cal R}e(\lambda_f) \sinh(y\tau)
-2 {\cal I}m(\lambda_f) \sin(x\tau) \Big\},\\
\label{tddr-2}
\Gamma(\overline{D}^0(t)\to f)&=&e^{-\tau}|\overline{A}_f|^2
\Big\{(1+|\lambda_f^{-1}|^{2}) \cosh(y\tau)
+(1-|\lambda_f^{-1}|^{2}) \cos(x\tau) \nonumber \\
&&+2 {\cal R}e(\lambda_f^{-1}) \sinh(y\tau)
-2 {\cal I}m(\lambda_f^{-1}) \sin(x\tau) \Big\}.
\eeqa
The time integrated rates are given by
\beqa\label{tddr-integrated}
\Gamma(D^0\to f)=\int_0^\infty\Gamma(D^0(t)\to f)\,dt &=&|A_f|^2
\Big\{(1+|\lambda_f|^2) {1 \over 1-y^2}
+(1-|\lambda_f|^2) {1 \over 1+x^2} \nonumber \\
&&+2 {\cal R}e(\lambda_f) {y \over 1-y^2}
-2 {\cal I}m(\lambda_f) {x \over 1+x^2} \Big\},\\
\label{tddr-integrated-2}
\Gamma(\overline{D}^0\to f)=\int_0^\infty\Gamma(\overline{D}^0(t)\to
f)\,dt&=&
|\overline{A}_f|^2
\Big\{(1+|\lambda_f^{-1}|^{2}){1 \over 1-y^2}
+(1-|\lambda_f^{-1}|^{2}) {1 \over 1+x^2} \nonumber \\ &&+2 {\cal
R}e(\lambda_f^{-1})  {y \over 1-y^2} -2 {\cal I}m(\lambda_f^{-1})
{x \over 1+x^2} \Big\}.
\eeqa
The corresponding expressions for decays into $\overline f$ follow via
the substitutions $f \to \overline f$ in the above expressions.

In general the four decay amplitudes can be written as
\beqa\label{fouramp}
A_f&=&A^T_{f} e^{+i\phi^T_{f}}[1+r_fe^{i(\delta_f+\phi_f)}],\qquad
A_{\overline f}=A^T_{\overline f} e^{i(\Delta_{f}+\phi^T_{\overline
f})} [1+r_{\overline f} e^{i(\delta_{\overline
f}+\phi_{\overline f})}] ,\nonumber\\
\overline{A}_{\overline f}&=&A^T_{f} 
e^{-i\phi^T_{f}}[1+r_fe^{i(\delta_f-\phi_f)}],\qquad
\overline{A}_f=A^T_{\overline f} e^{i(\Delta_{f}-\phi^T_{\overline f})}
[1+r_{\overline f }e^{i(\delta_{\overline f}-\phi_{\overline f })}].
\eeqa
where $A^T_{f} e^{\pm i\phi^T_{f}}$ is the Standard Model (SM) tree
level contribution. The phases $\phi_f^T$, $\phi^T_{\overline f}$,
$\phi_f$ and $\phi_{\overline f}$ are weak, CP violating phases, while
$\Delta_f$ and $\delta_f$ are strong, CP conserving phases. Neglecting
terms of order $|(V_{ub}V_{cb})/(V_{us}V_{cs})|\sim10^{-3}$,
$\phi^T_f=\phi^T_{\overline f}$ is the same for all final states.

%%%%%%%%%%%%
\subsection{CP eigenstates}
We consider final states that are CP eigenstates. (Note that this
analysis also applies to Cabibbo favored (CF) CP eigenstates, like
$K_S\pi^0$.) For a similar analysis see \cite{Du:2006jc}. For CP even
(odd) eigenstates, $\Delta_f=0$ ($\pi$). We can then write
\beqa\label{dircpv}
A_f&=&A^T_{f} e^{+i\phi^T_{f}}[1+r_fe^{i(\delta_f+\phi_f)}],\nonumber\\   
\eta^{\rm CP}_f \,\overline{A}_f&=&A^T_{f} 
e^{-i\phi^T_{f}}[1+r_fe^{i(\delta_f-\phi_f)}],
\eeqa
where $\eta^{\rm CP}_f=+(-)$ for CP even (odd) states.  Neglecting
$r_f$ in Eq. (\ref{dircpv}), $\lambda_f$ is universal and we can
define
\beq\label{lamind}
\lambda_f\equiv -\eta^{\rm CP}_fR_m\, e^{i \phi},
\eeq
where $R_m\equiv|q/p|$ and $\phi$ is the relative weak phase between
the mixing amplitude and the decay amplitude. The time-integrated CP
asymmetry for a final CP eigenstate $f$ is defined as follows:
\beq\label{akk}
a_f\equiv\frac
{\Gamma({D}^0\to f)-\Gamma(\overline{D}{}^0\to f)}
{\Gamma({D}^0\to f)+\Gamma(\overline{D}{}^0\to f)}.
\eeq

Given experimental constraints, we take $x,y,r_f\ll1$ and expand
to leading order in these parameters. Then, we can separate the
contributions to $a_f$ to three parts,  
\beq
a_f=a_f^d+a_f^m+a_f^i,
\eeq
with the following underlying mechanisms: 

(i) $a_f^d$ signals CP violation in decay:
\beq\label{afdir}
a^d_f = 2r_f \sin\phi_f \sin\delta_f. 
\eeq

(ii) $a_f^m$ signals CP violation in mixing. With our approximations,
it is universal:
\beq\label{afmix}
a^m=-\eta^{\rm CP}_f\,\frac{y}{2} (R_m-R_m^{-1}) \cos\phi.
\eeq

(iii) $a_f^i$ signals CP violation in the interference of decays with
and without mixing. With our approximations, it is universal:
\beq\label{afint}
a^i=\eta^{\rm CP}_f\,\frac{x}{2} (R_m+R_m^{-1}) \sin\phi .
\eeq

Consider the time dependent decay rates in Eqs. (\ref{tddr}) and
(\ref{tddr-2}). The
mixing processes modify the time dependence from a pure
exponential. However, given the small values of $x$ and $y$, the time
dependences can be recast, to a good approximation, into
purely exponential forms,
\beqa \label{time-dep}
\Gamma({D}^0(t)\to f)&\propto&\exp[-\hat\Gamma_{D^0\to f}\ t],\no\\
\Gamma(\overline{D}^0(t)\to f)&\propto&\exp[-\hat\Gamma_{\overline{D}^0\to
  f}\ t],
\eeqa
with modified decay rate parameters \cite{Bergmann:2000id}:
\beqa
\hat\Gamma_{D^0\to f}&=&
\Gamma_D[1+\eta^{\rm CP}_f\,R_m(y\cos\phi-x\sin\phi)],\no\\
\hat\Gamma_{\overline{D}^0\to f}&=&
\Gamma_D[1+\eta^{\rm CP}_f\,R_m^{-1}(y\cos\phi+x\sin\phi)].
\eeqa
One can define the following CP violating combination of these two
observables:
\beq\label{defdy}
\Delta Y_f\equiv\frac{\hat\Gamma_{\overline{D}^0\to f}-\hat\Gamma_{D^0\to f}}
{2\Gamma_D}=a^m+a^i.
\eeq

Note that $a^m$ and $a^i$ contribute to $a_f$ of Eq. (\ref{akk}) and
to $\Delta Y_f$ of Eq. (\ref{defdy}) in the same way, but $a_f^d$
contributes only to the former. In particular, $\Delta Y_f$ is
universal while $a_f$, in general, is not.

The experimental results from Babar \cite{Aubert:2003pz},
$\Delta Y=(-0.8\pm0.6\pm0.2)\times10^{-2}$,
and from Belle \cite{Abe:2003ys},
$\Delta Y=(+0.20\pm0.63\pm0.30)\times10^{-2}$, give the
following world average: 
\beq\label{dywa}
\Delta Y=(-0.35\pm0.47)\times10^{-2}.
\eeq

%%%%%%%%%%%%%%%%%%%%
\subsection{Non-CP eigenstates}
Here we consider final states that are not CP eigenstates.  For each
pair of CP conjugate states $f$ and $\overline f$, there are four
relevant amplitudes, Eq. (\ref{fouramp}). Neglecting $r_f$ and
$r_{\overline f}$ we have
\beq\label{defdf}
\lambda_f\equiv\frac qp\frac{\overline{A}_f}{A_f}
=-R_m R_f e^{i(\phi+\Delta_f)},\qquad
\lambda_{\overline f}\equiv\frac qp\frac{\overline{A}_{\overline
    f}}{A_{\overline f}} =-R_mR_f^{-1}e^{i(\phi-\Delta_f)}.
\eeq
Here $R_m$ and $\phi$ are the same as in Eq. (\ref{lamind}),
$R_f\equiv|\overline{A}_f/A_f|$, and $\Delta_f$ is
a strong (CP-conserving) phase. There are two time-integrated CP
asymmetries to consider:
\beq
a_f\equiv\frac {\Gamma({D}^0\to f)-\Gamma(\overline{D}^0\to\overline{f})}
{\Gamma({D}^0\to f)+\Gamma(\overline{D}^0\to\overline{f})},\qquad
a_{\overline{f}}\equiv\frac
{\Gamma({D}^0\to\overline{f})-\Gamma(\overline{D}^0\to{f})}
{\Gamma({D}^0\to\overline{f})+\Gamma(\overline{D}^0\to{f})}.
\eeq
Again, we take $x,y,r_f,r_{\overline f}\ll1$ and expand
to leading order in these parameters. Then
\beq \label{defamandai}
a_f=a_f^d+a_f^m+a_f^i,\qquad
a_{\overline f}=a_{\overline f}^d+a_{\overline f}^m+a_{\overline f}^i,
\eeq
where
\beqa \label{fin-non}
a^d_f &=& 2r_f \sin\phi_f \sin\delta_f, \no\\
a^m_f&=&-R_f\, \frac{y'_f}{2} (R_m-R_m^{-1}) \cos\phi, \no\\
a^i_f&=& R_f\, \frac{x'_f}{2} (R_m+R_m^{-1}) \sin\phi ,
\eeqa
(for $a_{\overline f}$ the result is the same with the replacement
$f \to {\overline f}$) and
\beqa
x^\prime_f&=&x\cos\Delta_f+y\sin\Delta_f,\ \ \ 
y^\prime_f=y\cos\Delta_f-x\sin\Delta_f,\no\\
x^{\prime}_{\overline f}&=&x\cos\Delta_f-y\sin\Delta_f,\ \ \ 
y^{\prime}_{\overline f}=y\cos\Delta_f+x\sin\Delta_f.
\eeqa
Since in SCS
decays we expect, in general, that $R_f={\cal O}(1)$, the decays into
final non-CP eigenstates should exhibit CP asymmetries of the same
order of magnitude as for CP eigenstates.

Several points are in order:
\begin{enumerate}
\item
One can, again, look for CP violation using the time dependence of
the decay, see Eq. (\ref{time-dep}). The result is similar to
Eq. (\ref{defdy}):
\beq\label{defdy-non}
\Delta Y_f\equiv\frac{\hat\Gamma_{\overline{D}^0\to
\overline f}-\hat\Gamma_{D^0\to f}}
{2\Gamma_D}=a^m_f+a^i_f, \qquad 
\Delta Y_{\overline f}\equiv\frac{\hat\Gamma_{\overline{D}^0\to
f}-\hat\Gamma_{D^0\to \overline f}}
{2\Gamma_D}=a^m_{\overline f}+a^i_{\overline f},
\eeq
where $a^m_f$ and $a^i_f$ are given in Eq. (\ref{fin-non}).
\item
A final state that is a CP eigenstates is a special case of the non-CP
final state, with $R_f=1$ and $\Delta_f=0$ ($\pi$)
for CP even (odd) final state. Then, Eqs. (\ref{fin-non})
reduce to Eqs. (\ref{afdir}), (\ref{afmix}) and (\ref{afint}).
\item
In analyses of CF and doubly Cabibbo suppressed (DCS) decays, such as
$D\to K\pi$, one usually finds expressions that depend on $x^\prime_f$
and $y^\prime_f$, but not on $x^{\prime}_{\overline f}$ and
$y^{\prime}_{\overline f}$ (see e.g. \cite{Kirkby:2004pu}). The reason
is not that the CP asymmetries are independent of
$x^{\prime}_{\overline f}$ and $y^{\prime}_{\overline f}$, but rather
that these contributions are relatively suppressed by
$\tan^4\theta_c$.
\end{enumerate}

%%%%%%%%%%%%%%%%
\subsection{Dalitz plot analysis for $D^0 \to VP$}
In practice, all final non-CP eigenstates are resonances. Thus, we can
perform a Dalitz plot analysis and sum up several resonances. Such an
analysis has several advantages. First, the statistics is
increased. Second, information about the strong phases can be
obtained.  A simple case is to concentrate on a single resonance in
the Dalitz plot, for example, $KK^*$. Then, from the interference
region of $K^+K^{-*}$ with $K^-K^{+*}$ the strong phase
$\Delta_{KK^*}$ can be determined \cite{Rosner:2003yk}.

The knowledge of the strong phase can be used to determine $x$ and
$y$, and not only $x^\prime_f$ and $y^\prime_f$. (Note that in the
standard analysis of DCS decays, only the latter can be determined.)
This can be seen by comparing the terms linear in $\tau$ to the
constant ones.  We see from Eqs. (\ref{tddr}) and (\ref{tddr-2}) that
we can measure the following four quantities:
\beq
y\, R_m \, \cos (\phi+\Delta_f),\quad
y\, R_m^{-1} \, \cos (\phi-\Delta_f),\quad
x\, R_m \, \sin (\phi-\Delta_f),\quad
x\, R_m^{-1} \, \sin (\phi+\Delta_f).
\eeq
Once these four quantities are measured, generally, one can separately
determine $x$, $y$, $R_m$ and $\phi$ (up to discrete ambiguities), and
thus separately measure the two types of indirect CP violation, $a^m $
and $a^i$. This cannot be done with a CP eigenstate.

%%%%%%%%%%%%%%%%%%%% 
\section{Direct vs. Indirect CP Violation}\label{sec:directvsin}
New CP violation could affect $a_f$ through either a contribution to
the mixing amplitude $M_{12}$, that is {\it indirect} CP violation, or
a contribution to the decay amplitudes $A_{f}$, that is {\it direct}
CP violation, or both. Indirect CP violation generates $a_f^m$ and
$a_f^i$, while direct CP violation generates $a_f^d$. (Contributions to
the decay amplitudes affect $\Gamma_{12}$ but this effect is always
very small and can be safely neglected.)

The SM contribution to the mixing is suppressed by three factors:
double Cabibbo suppression, flavor SU(3) suppression (which, in the
short distance language, is the GIM suppression) and weak-interaction
loop suppression. The long distance contribution avoids the loop
factor and can have a much milder SU(3)-breaking suppression.
Consequently, it is estimated that the SM gives $x,y={\cal
O}(10^{-3})$, but with very large uncertainties. In particular, it
cannot be excluded that the SM gives values as high as 
$x,y={\cal O}(10^{-2})$ \cite{Falk:2001hx,Falk:2004wg,Bigi:2000wn}.

New physics can avoid some or all of the three suppression
factors. Indeed, it is well known that there are many models that can
accommodate or even predict $x$ close to the current experimental
limit (for a review see \cite{Nelson:1999fg,Petrov:2003un}). The best
known example is that of supersymmetric models with quark-squark
alignment \cite{Nir:1993mx,Leurer:1993gy,Nir:2002ah}. Here, box
diagrams with intermediate squarks and gluinos have a double Cabibbo
suppression, but neither SU(3) nor $\alpha_w^2$-suppression (but only
$\alpha_s^2$ factor). Furthermore, the gluino couplings carry new CP
violating phases. These, and other models, demonstrate that it is
quite possible that indirect CP violation could account for $a_f$ of
${\cal O}(10^{-2})$.

Note that new short distance contributions can enhance $x$ but not
$y$. If the SM value of $y$ is small, $y \lsim 10^{-3}$, then
$a^m_f$  is negligible (in the case of a CP-eigenstate final
state). If $y$ is large, $y \sim 10^{-2}$, then new physics in the
mixing amplitude would result in similar contributions from $a_f^i$ and 
$a_f^m$.

The SM contribution to the decay is through tree level $W$-mediated
diagrams. Thus, the amplitude depends on $G_F\sin\theta_c$. New
physics cannot give competing contributions but, to generate
$a_f^d\sim10^{-2}$, it is only required that
\beq\label{satdir}
{\cal I}m(G_N)\sim10^{-2}\sin\theta_c\, G_F,
\eeq
where $G_N$ denotes the effective four-Fermi coupling from new
physics. If, for example, the scale of new physics is $\Lambda_{\rm
NP}\gsim1$ TeV then the scale-suppression of $G_N$ is ${\cal
O}(m_W^2/\Lambda_{\rm NP}^2)\lsim 10^{-2}$. Thus, quite generically,
Eq. (\ref{satdir}) can only be realized with $\Lambda_{\rm NP}\lsim1$
TeV and (at least) one of the following conditions satisfied: 
\begin{enumerate}
\item There is neither flavor-suppression stronger than $\sin\theta_c$
nor loop suppression;
\item There are enhancement factors related to hadronic factors or
chiral enhancement;
\item $\Lambda_{\rm NP}$ is actually much closer to $m_W$.
\end{enumerate}
As we show later, there exist well motivated models where indeed such
conditions apply and consequently (\ref{satdir}) can be satisfied. It
is thus quite possible that an ${\cal O}(10^{-2})$ effect is generated
solely or dominantly from direct CP violation.

In the absence of direct CP violation from new physics, the CP
asymmetries in SCS decays into final CP eigenstates would be {\it
  universal}, {\it i.e.} independent of the final state. (The SM would 
give tiny non-universal corrections, {\it i.e.}
$(a_{KK}-a_{\pi\pi})/(a_{KK}+a_{\pi\pi})={\cal O}
\{\arg[(V_{cd}^*V_{ud})/(V_{cs}^*V_{us})]\}\sim10^{-3}$.) 
We note that this universality would extend to CF decays to final CP
eigenstates, e.g., $D \to K_s \pi^0 $. Let us
define the universal, indirect contribution to CP violation as follows:
\beq
a^{\rm ind}= a^m+a^i.
\eeq
As mentioned above, $a^{\rm ind}$ is the {\it only}
possible source of $\Delta Y$ defined in Eq. (\ref{defdy}). Thus,
Eq. (\ref{dywa}) implies 
\beq
a^{\rm ind}=(-0.35\pm0.47)\times10^{-2}.
\eeq
We note that, if the time-integrated measurements yield a non-zero
asymmetry while the time-dependent measurements show no signal then
only direct CP violation must be playing a role.  More generally, if a
difference between the two classes of measurements is experimentally
established, and both are non-zero, then both direct and indirect CP
violation are present, and can be cleanly separated.  Such a scenario
is quite possible.  In fact, supersymmetric models with quark-squark
alignment
\cite{Nir:1993mx,Leurer:1993gy}
provide such an example, as we shall see.

We note that it is also possible to cleanly separate direct and
indirect CP violation in SCS decays only with time-integrated
CP asymmetry measurements.  Assuming negligible new CP violation
effects in CF and DCS decays (it is difficult to construct a model
in which this is not the case \cite{Bergmann:1999pm}), the
time-integrated CP asymmetry for a CF decay to a final CP eigenstate
would give the universal indirect CP asymmetry.  Subtracting this from
the time-integrated CP asymmetry for a SCS decay to a final CP
eigenstate would give the non-universal direct CP asymmetry for the
latter.  For example,
\begin{equation}
a_{P^+ P^-}^d  = a_{P^+  P^- } - a_{K_s \pi^0} \,,\qquad 
P = K,\,\pi  \,.
\end{equation}

Finally we mention that charged $D$ decays are sensitive only to
direct CP violation. If a non-vanishing CP asymmetry is experimentally
established in charged $D$ decay, that would signal direct CP
violation. If experiments establish time-integrated CP asymmetries in
neutral $D$ decays but not in charged $D$ decays, that would be
suggestive of indirect CP violation, but would not prove it.  Ii is
possible that the new physics could be such that it induces direct CP
violation only in neutral decays.

%%%%%%%%%%%%%%%%%%%%
\section{Direct CP violation at tree-Level}
\label{sec:directtree}
In this section we examine whether various specific models can
generate $a^d_f\gsim 10^{-2}$ via tree-level contributions. For
concreteness we focus on $f=K^+K^-$ and $\pi^+\pi^-$.
The main purpose is to find, for each model, an upper bound on the
$r_f$ factor of Eq. (\ref{dircpv}). We assume that the weak phase
$\phi_f$ is of ${\cal O}(1)$. The strong phase $\delta_f$ suffers from
hadronic uncertainties, but we point out cases where it is
formally suppressed by $1/N_c$. In practice, however, the strong phase
could be of ${\cal O}(1)$ even if it is color suppressed. 

%%%%%%%%%%%%
\subsection{Extra quarks in SM vector-like representations}
In models with non-sequential (`exotic') quarks, the $Z$-boson has
flavor changing couplings, leading to $Z$-mediated contributions to
the SCS decays. (For a review see, for example, \cite{Branco:1999fs}.)
In models with additional up quarks in the vector-like representation
$({\bf3},{\bf1},+2/3)\oplus({\bf\bar3},{\bf1},-2/3)$, the flavor
changing $Z$ couplings have the form
\beq
-{\cal L}_Z=\frac{gU_{ij}^u}{2\cos\theta_W}\ \bar u_{Li}\gamma_\mu
  u_{Lj}Z^\mu+{\rm h.c.} \quad \Longrightarrow\quad
G_N^Z=G_F U_{cu}^u .
\eeq
The flavor changing coupling is constrained by $\Delta m_D$
\cite{Bergmann:1999pm}: 
\beq
|U_{cu}^u|\lsim5\times10^{-4}\quad \Longrightarrow \quad 
r_f\lsim10^{-3}.
\eeq
A somewhat stronger bound (from $\Delta m_K$) applies for the case of
vector-like quark doublets, 
$({\bf3},{\bf2},+1/6)\oplus({\bf\bar3},{\bf2},-1/6)$.

We learn that a significant contribution to $D^0\to K^+K^-,\pi^+\pi^-$
from $Z$-mediated flavor changing interactions is ruled out. In fact,
this lesson applies to a much broader class of models, that is, all
models with a tree-level contribution mediated by a neutral heavy
boson. In all of these models, the combination $Y_{cu}/M$ (with
$Y_{cu}$ the flavor changing coupling and $M$ the mass of the heavy
boson) is constrained by $\Delta m_D$. The contribution to the decay
has an extra factor of $Y_{qq}/M$ ($q=s$ or $d$) that is maximized for
large $Y_{qq}$ and light $M$. Thus, the model discussed here, with
$Y_{qq}=g/(2\cos\theta_W)$ and $M=m_Z$, gives a contribution that is
near-maximal among all models with $Y_{qq}\lsim1$ and $M\gsim m_Z$.

%%%%%%%%%%%%
\subsection{Supersymmetry without $R$-parity}
We consider supersymmetry without $R$-parity models (for a description
of the framework, see, for example, \cite{Barbier:2004ez}).  The
lepton number violating terms in the superpotential
$\lambda^\prime_{ijk}L_i Q_j d_k^c$ give a slepton-mediated tree-level
contribution with an effective coupling
\beq
G_f^\prime=\frac{\lambda^\prime_{i2k}\lambda^{\prime*}_{i1k}}
{4\sqrt{2}M^2(\tilde\ell^-_{Li})}\quad {\rm with}\quad
k=\begin{cases}
  2&f=K^+K^-,\\
  1&f=\pi^+\pi^-.
  \end{cases}
\eeq
The same combinations of couplings contribute to the
$K^+\to\pi^+\nu\bar\nu$ decay. That provides the following bound (see
{\it e.g.} \cite{Barbier:2004ez}):
\beqa
|\lambda^\prime_{i2k}\lambda^\prime_{i1k}|
\times \left({100\ {\rm GeV}\over
M(\tilde d_k^c)}\right)^2 \lsim 2\times10^{-5}\quad
\Longrightarrow\quad
r_f\lsim1.5\times10^{-4},
\eeqa
where we take all sfermion masses to be of the same order.

The baryon number violating terms $\lambda^{\prime\prime}_{ijk}u_i^c
d_j^c d_k^c$ give a squark-mediated tree-level contribution with an
effective coupling
\beq
G_f^{\prime\prime}=\frac{\lambda^{\prime\prime}_{2jk}
\lambda^{\prime\prime*}_{1jk}}
{4\sqrt{2}M^2(\tilde d^c_{k})}\quad {\rm with}\quad
\begin{cases}
  j=2,k=1,3&f=K^+K^-,\\
  j=1,k=2,3&f=\pi^+\pi^-.
  \end{cases}
\eeq
Strong bounds are often quoted
from $n-\bar n$ oscillations (see
{\it e.g.} \cite{Barbier:2004ez}):
\beq
|\lambda^{\prime\prime}_{11k}|\lsim10^{-7}\quad ({\rm
  for}\ M(\tilde d_k^c)=100\ {\rm GeV}).
\eeq
(This would rule out any significant contribution to
$G^{\prime\prime}_{\pi\pi}$, and a significant contribution to
$G_{KK}^{\prime\prime}$ from $k=1$.)  However, it was shown in
\cite{Goity:1994dq} that important suppression factors were missed in
obtaining these bounds, and that the strongest individual bound on
these couplings comes from double nucleon decay,
\beq
|\lambda_{112}| < 10^{-15}\left({m_{\tilde g} m_{\tilde q}^4 \over  
\Lambda_h }\right)^{5\over 2}\,,
\eeq
where $\Lambda_h $ is some hadronic mass scale. This leaves only the
$k=3$ contributions to $G_{\pi\pi}$ and $G_{KK}$ as potentially 
significant (the revised bound from $n-\bar n$ oscillations in
\cite{Goity:1994dq}, $\lambda^{\prime\prime}_{113} < 0.002 (0.1)$ for
$m_{\tilde q} = 200\,(600)$ GeV, allows $r_{\pi\pi} \sim 10^{-2}$).
However, the $K^0-\overline{K}{}^0$ system yields the bounds
\cite{barbierimasiero,Barbier:2004ez} 
\beq
{\cal I}m(\lambda_{123}^{\prime\prime}\lambda_{113}^{\prime\prime
*})<10^{-5} \,,
~~~{\cal R}e(\lambda_{213}^{\prime\prime}\lambda_{223}^{\prime\prime
*})<3\cdot10^{-4}\,,~~~ 
{\cal I}m(\lambda_{213}^{\prime\prime}\lambda_{223}^{\prime\prime
*})<3\cdot10^{-6}\,, 
\eeq
from $\epsilon^\prime  /\epsilon$, $\Delta m_K$, and $\epsilon_K$,
respectively, for 100 GeV squark masses.
Note that each coupling appearing in these bounds also appears in
either $G_{\pi\pi}$ or $G_{KK}$, and {\it vice-versa}.
From this we conclude that it is not possible to simultaneously obtain
$r_{\pi\pi} \sim 10^{-2}$ and $r_{KK}\sim 10^{-2}$ for $k=3$, as this
would require a tuning among the $\lambda^{\prime\prime}$ couplings of
at least 1 part in $10^3$. (Also note that
$\lambda^{\prime\prime}_{ijk}\gsim10^{-7}$ would, in general, wash-out
a baryon asymmetry generated before the EWPT.)

In order to obtain a non-vanishing direct CP asymmetry in $D \to
K^+K^-$, a relative strong phase is required between the SM and NP
amplitudes. At the weak scale, the SM Hamiltonian mediating, {\it
e.g.}, $D\to K^+ K^- $ , is of the form $(\bar u_i s_i )_{V-A} (\bar
s_j c_j)_{V-A} $ ($i,j$ are color indices), while in the case of
R-parity violation, the relevant Hamiltonian is of the form
\beq
(\bar u_i s_i)_{V+A}(\bar s_j c_j)_{V+A}-(\bar u_i s_j)_{V+A}(\bar s_j
c_i)_{V+A}. \eeq 
Since the strong interactions conserve parity, the first term gives
the same strong phase as the SM. The second term, however, has a
different color structure and thus it can generate a different strong
phase. The contribution of the second term, however, is suppressed
compared to the first one by $1/N_c$. Thus, the
resulting strong phase relative to the SM amplitude is color
suppressed. As mentioned earlier, while this may mean that
the direct CP violation is further suppressed, an $O(1)$ relative
strong phase cannot be ruled out. The same argument applies to the $D
\to \pi^+ \pi^-$ amplitude.

%%%%%%%%%%%%
\subsection{Two Higgs doublet models (2HDM)}
We consider multi Higgs doublet models with natural flavor
conservation (for a review see, for example, \cite{Grossman:1994jb}).
In these models a charged Higgs ($H^\pm$) mediates a tree-level
contribution. In the 2HDM the relevant couplings are
\beq
-{\cal L}_{H^\pm}=\frac{ig}{\sqrt{2}m_W}\overline{u_i}
\left[m_{u_i}\cot\beta P_L+m_{d_j}\tan\beta P_R\right]  
V_{ij}d_jH^+ +{\rm h.c.}.
\eeq
It follows that the charged Higgs mediated contribution is also singly
Cabibbo suppressed. Then, for large $\tan\beta$, the suppression with
respect to the SM contribution is given by
\beq
r_{KK}\simeq
\frac{m_s^2\tan^2\beta}{m_{H^\pm}^2}.
\eeq
To obtain the upper bound, we consider the constraint on
$R_\tau\equiv{{\cal B}(B\to\tau\nu)}/{{\cal B}^{\rm
    SM}(B\to\tau\nu)}$ \cite{Isidori:2006pk}:
\beq
R_\tau\simeq
\left[1-\left(\frac{m_B}{m_{H^\pm}}\right)^2\tan^2\beta\right]^2\sim0.7\pm0.3. 
\eeq
We can write
\beq
r_{KK}\simeq
\frac{m_s^2}{m_b^2}\left(1-\sqrt{R_\tau}\right)\lsim 4\times10^{-4}.
\eeq
The bound on $r_{\pi\pi}^{H^\pm}$ is stronger by a factor of
$m_d^2/m_s^2$. For $\tan\beta\sim1$ the bound is even stronger,
$r_{KK}\simeq m_sm_c/m_{H^\pm}^2\lsim5\times10^{-5}$ (we use
\cite{PDG} $m_{H^\pm}\geq80$ GeV). We learn that the charged Higgs
contributions to the direct CP violation are negligible.

The situation is somewhat different in models with more than two Higgs
doublets. In particular, when two different doublets couple to the
down and charged lepton sectors, the bound from $B\to\tau\nu$ does not
apply to the SCS $D$ decays. One can still obtain a bound from charm
counting in $B$ decays. Using $n_{\rm charm}=1.22\pm0.04$, we conclude
that in this case $r_{KK}\lsim10^{-2}$ and
$r_{\pi\pi}\lsim10^{-4}$. Thus, direct CP violation from
charged Higgs contribution in 3HDM can marginally account for
$a_{KK}={\cal O}(10^{-2})$ but is negligible for $a_{\pi\pi}$.

%%%%%%%%%%%%
\section{Direct CP violation at one-loop}
\label{sec:directloop}
In the previous section we saw that, in models in which new decay
amplitudes are generated at the tree-level, the direct CP asymmetries
in SCS decays are typically constrained to lie well below the 1\%
level.  In this section we examine whether one-loop effects due to new
contributions to the $\Delta C=1$ QCD penguin and chromomagnetic
dipole operators can generate $a_f^d \sim 10^{-2}$.  Again, we
consider $KK$ and $\pi \pi$ final states, focus on $r_f $, and assume
that the new weak phase $\phi_f$ in Eq. (\ref{dircpv}) is of ${\cal
O}(1)$.

%%%%%%%%%%%%%%%%%%%%%%
\subsection{QCD penguin and dipole operators: General considerations}
\label{sec:genconsidns}
The $\Delta C=1$ effective Hamiltonian that is relevant to SCS decays
is given by
\begin{equation}
\label{Heff}
H_{\rm eff}^{\Delta C=1} = \frac{G_F}{\sqrt{2}} \left[\sum_{p=d,s} 
\lambda_p \left(C_1 Q_1^p + C_2 Q_2^p \right) + \sum_{i=3}^{6} 
C_i ( \mu ) Q_i ( \mu ) + C_{8g} Q_{8 g} \right]
+ {\rm H.c.},
\end{equation}
where $\lambda_p = V_{cp}^* V_{up}$ with $p = d,s$ are CKM factors,
and $\lambda_d + \lambda_s + \lambda_b=0$ due to the unitarity of the
CKM matrix. The operators are given in the appendix in
Eq. \ref{QCDoperators}).  $Q_{1,2}$ are the current-current operators,
$Q_{3,..,6}$ are the QCD penguin operators, and $Q_{8g}$ is the QCD
dipole operator.  The dominant contribution to the tree level
coefficients $C_1$ and $C_2$ is from the SM. New physics amplitudes
contribute to $C_{3,...,6}, C_{8g}$. The standard model contributions
to these operators can be neglected, as they enter at ${\cal O}(V_{cb}
V_{ub})$ (leading to $a_f^d \sim (V_{cb}V_{ub}/V_{cs}V_{us}
)\,\alpha_s /\pi \sim 10^{-4} $).  We have therefore opted to omit the
CKM factor in front of the penguin and dipole operators in
Eq. (\ref{Heff}). We emphasize that for CF decays, as well as DCS
decays, only the tree operators contribute. Penguin operators only
contribute to SCS decays.

There are also opposite chirality operators $\tilde Q_i$ which are
obtained from the $Q_i$'s via the substitutions $L\leftrightarrow
R$. In general their effects are of the same order of magnitude as the
operators that we discuss. In particular cases, like in left-right
symmetric models, there could be cancellations between the opposite
chirality contributions.  Here we consider only the general case where
such cancellations are not present. Furthermore, for simplicity we do
not write down explicitly the contributions of the opposite chirality
operators.

In many models the strongest bounds arise from $D^0-\overline{D}^0$
mixing.  The relevant $\Delta C=2$ effective Hamiltonian is given by
\cite{antichi}
\begin{equation}
\label{eq:eh}
H_{\rm eff}^{\Delta C = 2} = \sum_{i=1}^5 c_i O_i.
\end{equation}
Again, we do not write explicitly the opposite chirality operators
explicitly.  The operators $O_i$ are given in Eq. (\ref{eq:db2ops}) and
their matrix elements are estimated in Eq. (\ref{eq:dD2me}).
Experimental data yield bounds on the relevant operators. In
particular, we use \cite{Raz:2002ms}
\beq \label{DDbound}
|M_{12}^D| < 6.2 \times 10^{-11} \;{\rm MeV}.
\eeq

In order to obtain rough estimates of the $D \to KK$ and
$D\to\pi\pi$ amplitudes we use the QCD factorization framework
\cite{Beneke:2001ev}. 
We adapt the original $B$ decay discussion of \cite{Beneke:2001ev} to
the case of $D$ decays. We work primarily at leading order in $1/m_c$,
using naive factorization for $Q_{1,..,6}$, and QCD factorization for
$Q_{8g}$. We identify, however, possibly large power corrections
associated with the annihilation topology for the current-current and
penguin operators, which formally enter at ${\cal O}(1/m_c)$.

Clearly, the $1/m_c$ expansion is not expected to work very well for
hadronic $D$ decays.  Thus, our analysis only provides order of
magnitude estimates for the full decay amplitudes, which suffice for
our purposes.  It should also be noted that the QCD factorization
approach is useful for organizing the matrix elements of the various
operators in order of importance.  

In Appendix A we give the details of our analysis and quantitative
estimates.  Our conclusions with regard to annihilation amplitudes can
however be simply stated:
\begin{itemize}
\item For the SM operators, the spectator and the annihilation
amplitudes are roughly of the same order (see Eq. (\ref{AKTK}) for
details).
\item For the penguin operators, the annihilation amplitudes
are likely to give the dominant contribution (see Eq. (\ref{AKTK-NP})
for details). 
\end{itemize}

%%%%%%%%%%%%%%%%%%%%%%
\subsection{Implications of Isospin and $SU(3)_F$}

Model independently there are no significant bounds on the relevant
operators, so we can get $a_f^d\sim 10^{-2}$. There are, however,
several general results that can be obtained based on symmetries, in
particular, isospin and U-spin.

Very generally isospin predicts 
\beq
A(D^0  \to  \pi^0 \pi^0 )+ \sqrt{2}\,{A}(D^+ \to \pi^+ \pi^0) 
 -A(D^0 \to \pi^+
\pi^-)=0. 
\eeq
As for the new penguin amplitudes, the isospin predictions
follow from the fact that the $c\to ug$ operator is $\Delta
I=1/2$. Thus, it cannot generate an $I=2$ final state. In particular,
it cannot contribute to $D^+ \to \pi^+\pi^0$. Thus, we expect no
direct CPV in this mode, $a_{\pi^+\pi^0}=0$.  In contrast, we can get
direct CPV in $D^0$ decays as well as in $D^+\to K^+K_S$.  Other
isospin-based predictions would need further assumptions. For example,
neglecting annihilation diagrams, isospin predicts that
$a_{K^+K^-}^d=a_{K^+K_S}^d$.  As we just argued, neglecting
annihilation cannot be justified.  In principle, it could flip the
sign between the two asymmetries.

U-spin predicts that $a^d_{K^+K^-}=-a^d_{\pi^+\pi^-}$ for new $c \to u
g$ transitions. (This is in contrast to the indirect CP violation
which gives the same sign, $a^{\rm ind}_{K^+K^-}=a^{\rm
ind}_{\pi^+\pi^-}$.)  U-spin predicts that the SM amplitudes for the
two processes have opposite signs ($O(\lambda^4)$ effects coming from
${(V_{cs} V_{us}^*)/(V_{cd} V_{ud}^*)} \ne 1$ are negligible), whereas
penguin amplitudes have the same sign. Further study of U-spin
violation, especially in annihilation, is needed in order to check the
resulting prediction of opposite signs for $a^d_{K^+K^-}$ and
$a^d_{\pi^+\pi^-}$.

Another U-spin prediction is that in the SM $A(D\to K^0
\overline{K}^0)$ vanishes. This is a pure annihilation process with two
contributing diagrams: One where $c\bar u\to d\bar d$ ($\propto V_{cd}
V_{ud}^*$) and the $s \bar s$ pair pops out of the vacuum, and a
second one where $c\bar u\to s\bar s$ ($\propto V_{cs} V_{us}^*$) and
the $d\bar d$ pair pops out of the vacuum. Again, due to the sign
difference between the two CKM combinations, the total amplitude is
proportional to $d \bar d - s \bar s$ which vanishes in the U-spin
limit.  Thus, the data (\ref{KKBRs}) shows not only that annihilation
is large but also that U-spin breaking is large for annihilation.

%%%%%%%%%%%%%%%%%%%%%%%%%%%%%%%
\subsection{QCD penguin and dipole operators:  Examples from SUSY}
\label{sec:susyegs}
We study contributions to the QCD penguin and dipole operator Wilson
coefficients arising from up squark-gluino loops.  For simplicity, we
work in the squark mass-insertion approximation. The common squark
mass is denoted by $\tilde{m}$. We consider the contributions of the
up-squark mass insertions
\beq
\delta_{LL} \equiv \frac{(\tilde{m}^{2u}_{LL})_{12}}{\tilde{m}^2} , \qquad
\delta_{LR} \equiv \frac{(\tilde{m}^{2u}_{LR})_{12}}{\tilde{m}^2}.
\eeq
(The opposite chirality mass insertions $\delta_{RR}$ and
$\delta_{RL}$ are obtained via the substitutions $L\leftrightarrow R$
above.)  The Wilson coefficients are given by 
\beq
C_{i} =  E_{i} (x)\, \delta_{LL},~~i=3,..,6, \qquad 
C_{8g} = F(x)\, \delta_{LL} + G(x)\, \frac{m_{\tilde g}}{m_c} \,
\delta_{LR}\,,
\eeq
where $x= m_{\tilde{g}}^2 /\tilde m^2 $.  $E_i (x)$, $F(x)$, and
$G(x)$ contain loop functions, and can be read from
Eq. (\ref{susyCidecay}).  We learn that $\delta_{LL}$ contributes to
all of the penguin operators, while $\delta_{LR}$ only contributes to
$C_{8g}$.  Note that the contribution from $\delta_{LR}$ is enhanced
by a large factor of $m_{\tilde g}/m_c $. In addition, the loop
function $G(x)$ that accompanies $\delta_{LR}$ gives a further
enhancement, which is numerically of order five in the relevant
parameter space, relative to $F(x)$.

The most severe bounds arise from $D^0 - \overline{D}{}^0$ mixing.
The full expressions for the Wilson coefficients are given in
Eq. (\ref{dD2wilsonSUSY}).  What we find is that all of the mass
insertions enter the expressions with similar coefficients. In
particular, there is no enhancement for the chirality changing
insertions.

We begin with a discussion of the effects of the left-right squark
mass insertion, $\delta_{LR}$.  It generates new contributions to
$D$-meson decays via the dipole operator $Q_{8g}$, and to
$D^0-\overline{D}{}^0$ mixing via the operators $O_{2},\, O_3$. The
crucial point is that the contribution to the decay (but not to the
mixing) is enhanced by a large factor, $m_{\tilde g} / m_c$, and
therefore the $D^0-\overline{D}{}^0$ mixing bounds
are not restrictive. Consequently, $O(10^{-2})$ contributions to the
$D\to KK,\pi\pi$ amplitudes are not excluded. 

\begin{figure}
\begin{center}
\hbox{$\!\!\!\!\!\!\!\!\!\!\!\!\!\!\!\!\!\begin{array}{cc}
\includegraphics[width=9.5cm]{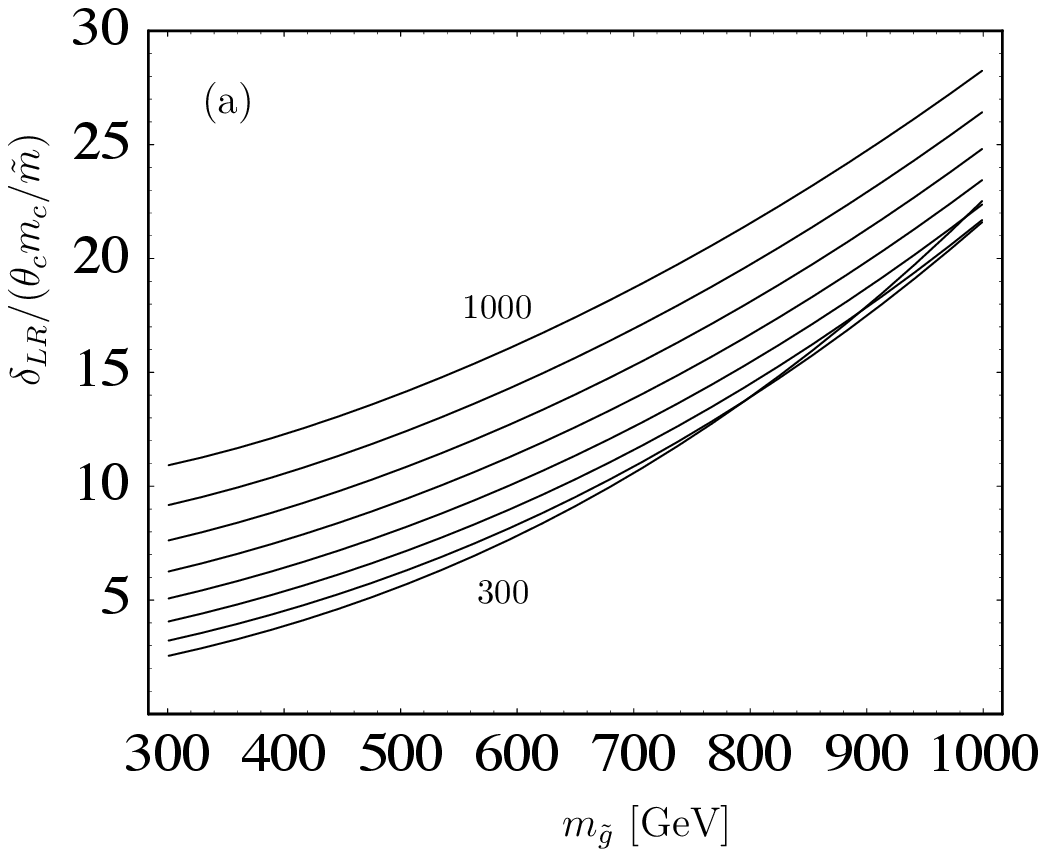} \!\!&\!\!
\includegraphics[width=9.5cm]{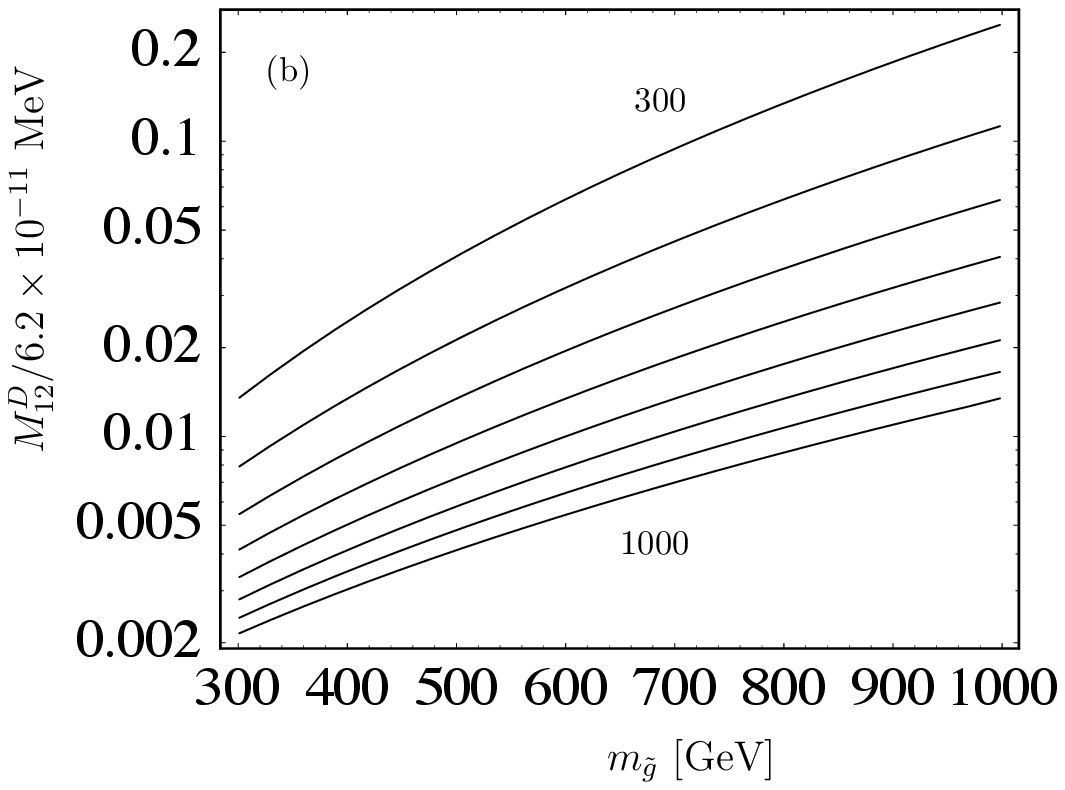}
\end{array}$}
\end{center}
\centerline{\parbox{14cm}{\caption{\label{fig:deltaLR}    
(a) $\delta_{LR}$ [in units of $(\theta_c m_c /\tilde m)$] {\it vs.}
$m_{\tilde g}$, and (b) $M_{12}^D$ [in units of $6.2\times 10^{-11}$
MeV] {\it vs.} $m_{\tilde g}$, for $r_f=0.01$ ($f=K^+K^-,\pi^+\pi^-$).
The lines correspond to $\tilde{m} = 300-1000$ GeV in increments of
100 GeV.}}} 
\end{figure}

The situation is illustrated in Fig.~\ref{fig:deltaLR}. The
contours in these plots correspond to a fixed ratio, $r_f=
10^{-2}$. This ratio is calculated using QCD factorization at
leading-power for the dipole operator amplitude and naive
factorization for the standard model amplitude, see Eqs. (\ref{ai})
and (\ref{leadingpoweramps}).  In Fig.~\ref{fig:deltaLR}(a) we plot
the values of $\delta_{LR}$ that yield $r_f=10^{-2}$ as a function of
the gluino mass, $m_{\tilde g}$, for several values of $\tilde{m}$.
$\delta_{LR}$ is plotted in units of $\theta_c m_c (\mu_{\rm
susy})/\tilde m$. (For simplicity, we take $\mu_{\rm susy}=m_t $ and
neglect the small running of $m_c$ between $m_t$ and the squark mass
scale, which yields $m_c(\mu_{\rm susy})=0.85$ GeV for $m_c(m_c)=
1.64$ GeV.) This is useful for later comparison to the magnitudes
expected for $\delta_{LR}$ in various supersymmetric models of flavor.
In Fig.~\ref{fig:deltaLR}(b) we plot the corresponding contributions
to $|M_{12}^D|$, normalized to the upper bound of $6.2\times 10^{-11}$
MeV, see Eq. (\ref{DDbound}). We learn that it is possible to obtain
$O(10^{-2})$ contributions to the decay amplitudes, accompanied by new
contributions to $|M_{12}^D|$ lying one to two orders of magnitude
below the experimental bound.  In the standard model the annihilation
amplitude could be of same order as the leading power tree amplitude
with large relative strong phase (this is probably also true for the
annihilation vs. leading power dipole operator amplitudes).
Therefore, if $\arg(\delta_{LR})$ is large, then $a_f^d=O(10^{-2})$
could be realized with negligible $a_f^{\rm ind}$.  A striking feature
of this result is the sensitivity of current CP asymmetry searches to
very small values of $\Im(\delta_{LR}) \gsim 2 \times 10^{-3}$.

\begin{figure}
\begin{center}
\includegraphics[width=9.5cm]{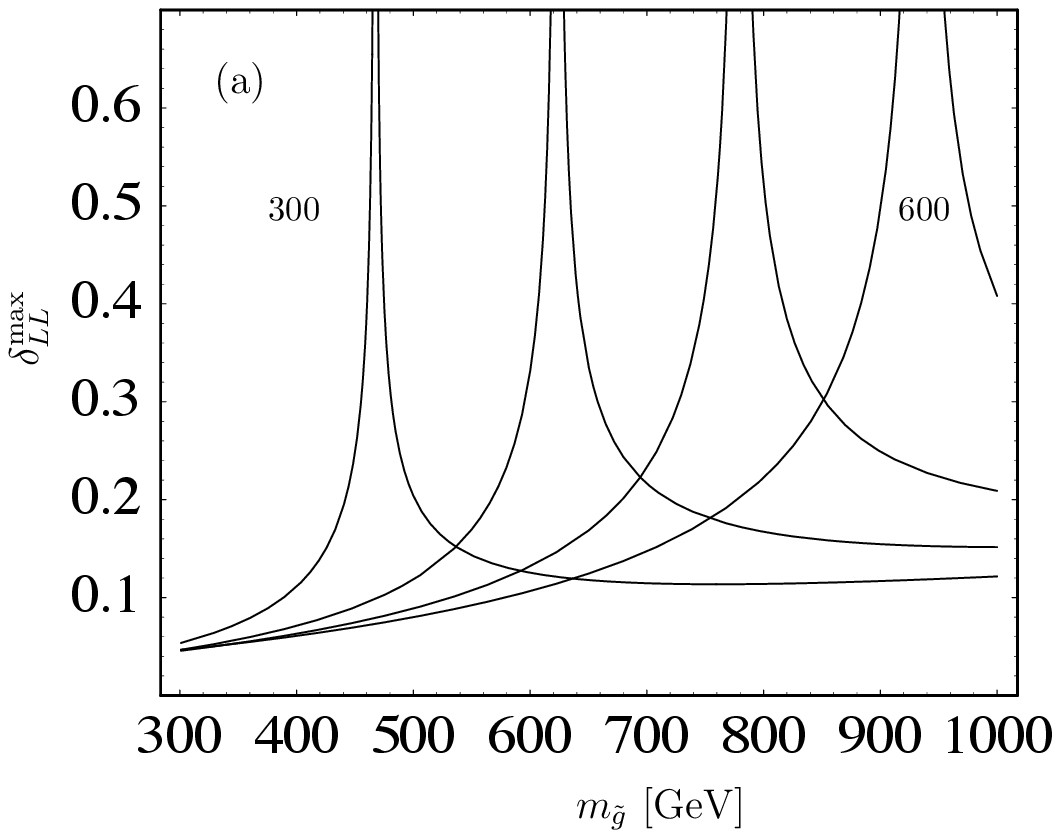} \\
\hbox{$\!\!\!\!\!\!\!\!\!\!\!\!\!\!\!\!\!\begin{array}{cc} 
\includegraphics[width=9.5cm]{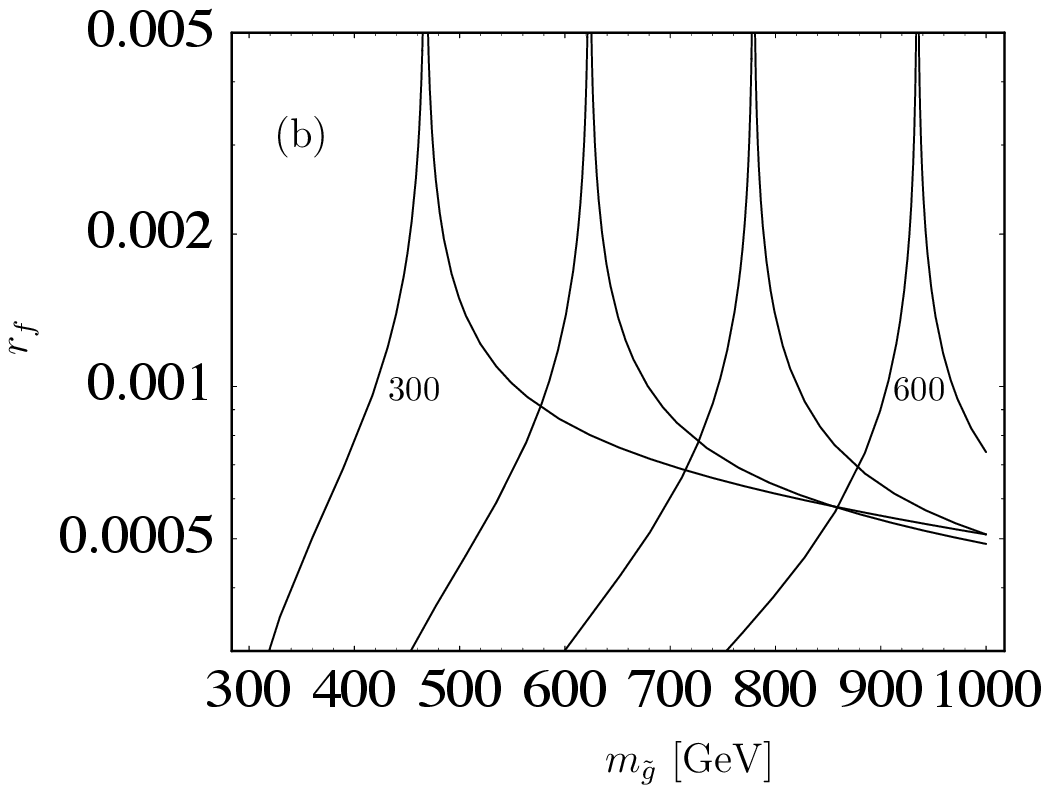} \!\!\!&\!\!\! 
\includegraphics[width=9.5cm]{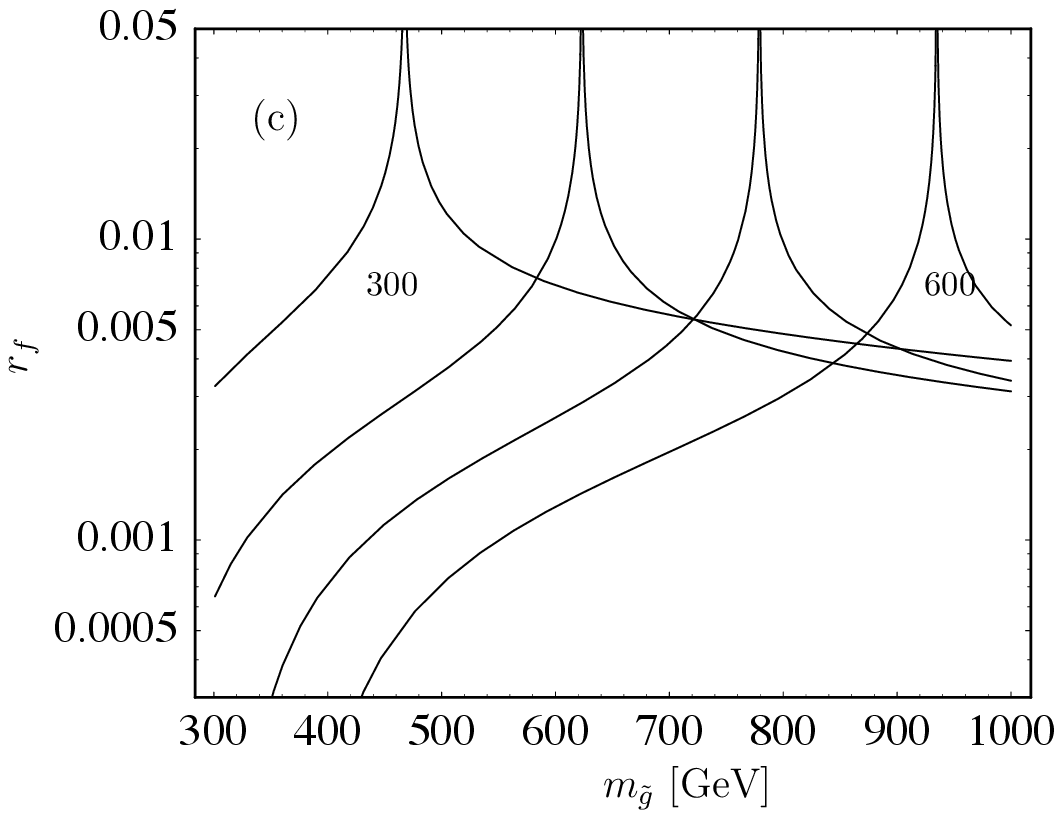} 
\end{array}$} 
\end{center} 
\centerline{\parbox{14cm}{\caption{\label{fig:deltaLLfig}    
(a) $\delta_{LL}^{\rm max}$ (the upper bound on $\delta_{LL}$ from
$D^0-\overline{D}{}^0$ mixing) {\it vs.} $m_{\tilde g}$; (b,c) $r_f$
($f= K^+ K^- ,\,\pi^+ \pi^-$) corresponding to $\delta_{LL}^{\rm max}$
{\it vs.} $m_{\tilde g}$ (b) in naive factorization or (c) with
annihilation power corrections included in $A^{NP}$ (see text). The
various lines correspond to $\tilde m = 300,\,400,\,500,\,600$ GeV.}}}
\end{figure}

Next, we discuss the effects of the left-left squark mass insertion,
$\delta_{LL}$.  New contributions to the $D$ decay amplitudes are
generated via the QCD penguin and dipole operators $Q_{3,..,6}$,
$Q_{8g}$.  Their magnitudes are restricted by requiring that the
supersymmetric contribution to $|M_{12}^D|$ is smaller than the bound
in Eq. (\ref{DDbound}). Here, unlike in the case of $\delta_{LR}$,
there is no $m_{\tilde g} /m_c $ enhancement of the contribution to
the decay and, consequently, the bound from the mixing is
significant. In Fig.~\ref{fig:deltaLLfig}(a) the resulting upper bound
on $\delta_{LL}$ is plotted as a function of $m_{\tilde g}$ for
several values of $\tilde m$.  The corresponding upper bounds on $r_f$
($f=K^+K^-,\pi^+\pi^-$) are plotted in
Fig.~\ref{fig:deltaLLfig}(b). (Again, the hadronic matrix elements of
the four-quark operators and the dipole operator are estimated in
naive factorization and in QCD factorization, respectively.) The
supersymmetric contribution to $M_{12}^D$ in Eq. (\ref{dD2wilsonSUSY})
vanishes at $m_{\tilde g}\approx 1.56 \tilde m $ leading to the peaked
structures in Fig.~\ref{fig:deltaLLfig}, also see \cite{Nir:2002ah}.
In the absence of special tuning of $m_{\tilde g}$ vs. $\tilde m$, we
observe that at leading-power $r_f \lsim 10^{-3}$.  (We note that the
validity of the squark mass-insertion approximation is marginal for
$\delta_{LL} \gsim 1/4 $, but it is sufficient for our purposes given
the much larger hadronic theoretical uncertainties \cite{Raz:2002zx}).

It may well be the case, however, that the $1/m_c$ expansion fails
badly in the evaluation of the QCD penguin contributions. In
particular, as argued in Appendix A, annihilation amplitudes could
give an order of magnitude enhancement. To show how the situation
changes if such enhancement is indeed realized, we repeat the
calculation with QCD penguin annihilation matrix elements included
according to Eqs. (\ref{Annamps}), (\ref{Ai}) and (\ref{Aibi}).  As
discussed in Appendix A, we estimate these matrix elements in the
one-gluon exchange model of \cite{Beneke:2001ev,Beneke:2003zv}.  
The results are presented in Fig.~\ref{fig:deltaLLfig}(c). Our
conclusion is that, if annihilation enhances the QCD penguin operator
contributions, then it is possible that supersymmetric $\delta_{LL}$
insertions give $a_f^d\sim10^{-2}$ without violating the bounds from
mixing.  In other words, due to hadronic uncertainties, we cannot rule
out the possibility of such large direct CP violation from
$\delta_{LL}$. In this case, however, we also expect the indirect CP
violation to be of same order.

%%%%%%%%%%%%%%%%%%%%%%
\subsection{FCNCs in supersymmetric flavor models}
Supersymmetric models with minimal flavor violation, such as gauge or
anomaly mediation, give no observable CP violating effects in SCS $D$
decays. We thus consider supersymmetric models where the SUSY breaking
mediation is not flavor blind. In such models there are two main
strategies for suppressing FCNCs: (a) quark-squark alignment
\cite{Nir:1993mx,Leurer:1993gy,Nir:2002ah}, (b) squark mass
degeneracy, see e.g.,
\cite{Dine:1993np,Pomarol:1995xc,Barbieri:1995uv,Barbieri:1996ae,Barbieri:1996ww,Carone:1996nd,Barbieri:1998em,Masiero:2001cc}.
Models in each category make specific predictions for the pattern of
squark mixing, or for the squark mass-insertions $\delta_{NM}$,
($N,M=L,R$).  In the following, we compare these predictions with the
sensitivity of current direct and indirect CP asymmetry searches.

The various models are based on approximate horizontal symmetries, and
often make predictions in terms of a small symmetry breaking
parameter. For concreteness, we use
$\lambda\sim\sin\theta_c\sim0.2$ as the small parameter.

In models of alignment, Abelian flavor symmetries are responsible for
the observed quark mass and mixing hierarchies and lead to a high 
degree of alignment between the down quark and down squark mass
eigenstates. Thus, supersymmetric FCNCs in the down sector are highly 
suppressed. CKM mixing is generated in the up sector, and the up
squarks are non-degenerate. The models make the following order of 
magnitude predictions \cite{Nir:2002ah}:
\beq\label{aligndel}
\delta_{LR} \sim {\lambda m_c \over \tilde{m} },\qquad
\delta_{LL} \sim \lambda, \qquad
\delta_{RR} \lsim \lambda^2, \qquad 
\delta_{RL} \lsim {\lambda^2 m_c\over\tilde{m}}.
\eeq
In addition, ${\cal O}(1)$ CP violating phases are expected. 

Comparing the predicted range for $\delta_{LR}$, Eq. (\ref{aligndel}),
with the values required to generate $r_f\sim0.01$,
Fig.~\ref{fig:deltaLR}(a), one may naively conclude that alignment
gives values of $r_f$ that are a factor of $3-30$ too small. It should
be kept in mind, however, that the dipole operator matrix elements
suffer from large theoretical uncertainties.  In particular, we have
not taken into account power corrections due to the annihilation
topology.  Therefore, an enhancement of $r_f$ by a factor of a few
cannot be ruled out. We conclude that for squark and gluino masses at
the lower part of the range that we consider, $\delta_{LR}$ could lead
to $a_f^d \sim 10^{-2}$.  According to Fig.~\ref{fig:deltaLR}(b), the
contribution of $\delta_{LR}$ to indirect CP violation is bounded
to be small.

Comparing the predicted range for $\delta_{LL}$, Eq. (\ref{aligndel}),
with the values required to generate $r_f\sim0.01$, 
Figs.~\ref{fig:deltaLLfig}(b) and \ref{fig:deltaLLfig}(c), we learn that
$\delta_{LL}$  could also lead to $a_f^d \lsim 10^{-2}$,
provided that annihilation strongly enhances the penguin operator
matrix elements.  Finally, Fig.~\ref{fig:deltaLLfig}(a) confirms that the
predicted range for $\delta_{LL}$ could easily lead to $a_f^i \sim
10^{-2}$ and, if $y \sim 10^{-2}$, also to $a_f^m \sim 10^{-2}$. We
conclude that models of alignment predict $a^{\rm ind} \sim 10^{-2}$
and could also accommodate $a^d \lsim 10^{-2}$.

In models of squark degeneracy, the first
two families of quarks constitute a doublet, and the third family a
singlet, of a non-Abelian horizontal symmetry.  This
leads to a high degree of degeneracy between the first and second
family squark masses which evades the bounds from $\Delta m_K$, and
implies $\delta_{LL},\,\delta_{RR} \ll 1$.  Thus, the contributions of
$\delta_{LL}$ and $\delta_{RR}$ to $a_f^d $ and $a_f^{\rm ind}$ are
negligible.  The non-Abelian horizontal symmetry is not sufficient for
reproducing all features of the quark mass and mixing hierarchies
without a large Yukawa coupling hierarchy, and may not lead to a
sufficiently high degree of degeneracy between the down and strange
squark masses to evade the bounds from $\epsilon_K$.  Thus, an Abelian
flavor symmetry is introduced (it could be a subgroup of a larger
non-Abelian symmetry). The resulting predictions for $\delta_{LR}$ and
$\delta_{RL}$ are model-dependent.  For example, $U(2)$ based models,
with vanishing $(1,1)$ entries in the quark mass matrices
\cite{Barbieri:1995uv,Barbieri:1996ae,Barbieri:1996ww,Barbieri:1998em,Masiero:2001cc},
predict 
\beq
\delta_{LR} \sim \delta_{RL} \sim \frac{\sqrt{m_u m_c}}{\tilde m} 
\sim \frac{\lambda^2 m_c}{\tilde m}.
\eeq
Therefore, in such models the contributions of $\delta_{LR}$ and
$\delta_{RL}$ to $a_f^d$ are well below $10^{-2}$.  
The effect can be larger in models with a discrete non-Abelian 
$S_3^3$ horizontal symmetry \cite{Carone:1996nd},
which predict $\delta_{LR}\sim {\lambda m_c}/{\tilde m}$ and 
$\delta_{RL} \sim {\lambda^3 m_c}/{\tilde m}$, quite similar to
models of alignment. Therefore, $a_f^d \sim 10^{-2}$ may again be
possible via the dipole operator.  

We conclude that $a_f^d \sim 10^{-2}$ is not generic but could arise
in specific models of squark degeneracy via the dipole operator with
negligible mixing effects. In models of alignment, $a_f^d\sim10^{-2}$
can arise via the dipole operator as well as the penguin operators,
the latter being correlated with a large mixing contribution that is
likely to yield $a_f^{\rm ind} \sim 10^{-2}$.  In both examples a
significant dipole operator contribution to $a^d_f$ is linked to a
large contribution to $\theta_c$ from the up quark sector.

It is interesting to compare the sensitivity of CP violation in SCS
$D$ decays and in $B$ decays to models of flavor. The two sectors
provide complementary information. The combination of measurements of
$D$, $B_d$, $B^+$ and $B_s$ decays can be used to discriminate between
different models of flavor. The details of the comparison are left for
a future publication.

%%%%%%%%%%%%%%%%%%%%%%%%%%%%%%%%%%%%%%%%
\section{Discussions and conclusions}
\label{sec:conclusions}
It is well known that CP violation in $D$ decays is a clean way to
probe new physics. In this paper we study CP asymmetries in singly
Cabibbo suppressed (SCS) $D$ decays, focusing in particular on the
final CP eigenstates $K^+K^-$ and $\pi^+\pi^-$. The possibility to
probe new CP violation is, however, not limited to these
modes. Pseudo-two body CP eigenstates, such as $\phi \pi^0$ or $\phi
K_S$, as well as non-CP eigenstates, for example $KK^*$ and $\rho \pi
$, are also worth studying.  In particular, we have seen that the
formalism for time-integrated CP asymmetries in decays to non-CP
eigenstates allows a separation of indirect CP violation due to mixing
and due to interference of decays with and without mixing.  Decays
with four (or more) final state particles, like $\rho^0 \rho^0$, offer
new ways to probe CP violation via triple product correlations. It is
likely that models that lead to large direct CP asymmetries in two
body decays also generate large CP violating triple products.

To summarize, our main results are as follows:
\begin{itemize}
\item 
The SM cannot account for asymmetries that are significantly larger
than ${\cal O}(10^{-4})$. Thus, CP violation from new physics must be
playing a role if an asymmetry is observed with present experimental
sensitivities [${\cal O}(0.01)$].
\item 
The underlying mechanism of CP violation can be any of the
three types: in decay ($a^d$), in mixing ($a^m$), and in the
interference of decays with and without mixing ($a^i$).
\item 
In the case of indirect CP violation ($a^{\rm ind}=a^m+a^i$) and final
CP eigenstates, the time integrated CP asymmetries $a_f$ and the time
dependent asymmetries $\Delta Y_f$ are universal (and equal to each
other).  
\item
In contrast, for direct CP violation, the time integrated asymmetries
$a_f$ are not expected to be universal, while the time dependent
asymmetries $\Delta Y_f$ vanish. 
\item 
The pattern of CP violation can be used to test supersymmetric flavor 
models. Minimal flavor violation models predict tiny, unobservable,
effects. Alignment models predict large $a^{\rm ind}$ and possibly 
also large $a_f^d$. Models with squark degeneracy predict small $a^{\rm
ind}$ but, depending on the model, can accommodate observable $a_f^d$.
\item 
If direct CP violation is at the $1\%$ level, its likely source is 
new physics that contributes to the decay via loop diagrams
rather than via tree diagrams. The reason is that the experimental
bounds on $D^0-\overline{D}{}^0$ mixing are much more effective in
constraining models of the latter type.
\item
In this regard, SCS $D$ decays are unique, as they are the only ones
that probe gluonic penguin operators. In other words, while we find
that direct CP violation can have observable effects in SCS decays, it
is very unlikely to affect CF and DCS decays. 
\end{itemize}

%%%%%%%%%%%%%%%%%%%%%%%%%%%%%
\section*{Acknowledgments} 
We are grateful to Brian Meadows, Kalanand Mishra, and Mike Sokolof
for useful discussions.  A.K. would like to thank the Technion and
Weizmann Institute Physics Departments for their hospitality
throughout the course of this work.  This project was supported by the
Albert Einstein Minerva Center for Theoretical Physics, and by EEC RTN
contract HPRN-CT-00292-2002. The work of Y.G. is supported in part by
the Israel Science Foundation under Grant No. 378/05. The research of
A.K. is supported in part by the U.S. Department of Energy, under
grant DOE-FG02-84ER-40153, and by a Lady Davis Fellowship.  The
research of Y.N. is supported by the Israel Science Foundation founded
by the Israel Academy of Sciences and Humanities, and by a grant from
the United States-Israel Binational Science Foundation (BSF),
Jerusalem, Israel.

%%%%%%%%%%%%%%%%%%%%%%%%%%%%%
\appendix
\section{The $D\to KK/\pi\pi$ amplitudes}
We use the QCD factorization framework \cite{Beneke:2001ev} to obtain
order of magnitude estimates for the $D \to KK/\pi\pi$ amplitudes in
the presence of new contributions to the QCD penguin and dipole
operators. Clearly, the $1/m_c$ expansion is not expected to work very
well for hadronic $D$ decays. We can therefore ignore ${\cal
O}(\alpha_s) $ corrections to the matrix elements, as they are
negligible compared to the overall theoretical uncertainties.  We work
primarily at leading order in $\Lambda_{\rm QCD}/m_c$, using naive
factorization for $Q_{1,..,6}$ and QCD factorization for $Q_{8g}$.
However, we discuss the importance of power corrections, especially
annihilation, in the standard model and estimate a large source of
theoretical uncertainty in the QCD penguin operator matrix elements
due to annihilation.

Our convention for the flavor wave functions is
\beqa
\label{flavorconvention}
\pi^0 &\sim & {1\over \sqrt{2} } (\bar u u -\bar d d)\,,~~~\pi^- \sim
\bar u d \,~~~\pi^+ \sim \bar d u \nonumber\\
K^0 &\sim& \bar d s \,,~~K^0 \sim \bar s d \,,~~K^- \sim \bar u
s\,,~~K^+  \sim \bar s u\,.
\eeqa

%%%%%%%%%%%%%%%%%%%%%%%%%%%%%%%%%% 
\subsection{Leading-power}
\label{sec:leadingpower}

The effective $\Delta C=1$ Hamiltonian is given in Eq. (\ref{Heff})
\begin{equation}
\label{Heff-afain}
H_{\rm eff}^{\Delta C=1} = \frac{G_F}{\sqrt{2}} \left[\sum_{p=d,s} 
\lambda_p \left(C_1 Q_1^p + C_2 Q_2^p \right) + \sum_{i=3}^{6} 
C_i ( \mu ) Q_i ( \mu ) + C_{8g} Q_{8 g} \right]
+ {\rm H.c.},
\end{equation}
The operators are given by:
\beq
\label{QCDoperators}
\begin{array}{lcl}
Q_1^p=(\bar p c)_{V-A}(\bar u p)_{V-A} & &
Q_2^p=(\bar p_\alpha c_\beta)_{V-A} (\bar u_\beta
p_\alpha)_{V-A} \nonumber \\ Q_3=(\bar u c)_{V-A} \sum_q (\bar
qq)_{V-A} && Q_4=(\bar u_\alpha c_\beta)_{V-A} \sum_q (\bar q_\beta
q_\alpha)_{V-A} \nonumber \\ Q_5=(\bar u c)_{V-A}\sum_q(\bar qq)_{V+A}
&& Q_6=(\bar u_\alpha c_\beta)_{V-A}\sum_q(\bar q_\beta
q_\alpha)_{V+A}
\nonumber \\
Q_{8g} = -\frac{g_s}{8\pi^2}\, m_c \bar u \,\sigma_{\mu\nu}(1+\gamma_5)
G^{\mu\nu} c &&
\end{array}
\eeq
where $\alpha,\beta$ are color indices and $q=u,d,s$.
The matrix elements for $D\to KK,\,\pi\pi$
decay can be written in the form \cite{Beneke:2001ev,Beneke:2003zv}
\begin{equation}\label{Top}
   \langle P_1  P_2 |{\cal H}_{\rm eff}| D \rangle
   = \langle P_1 P_2 |{\cal T}_A+{\cal T}_B |D\rangle \,,
\end{equation}
where ${\cal T}_A$ is the transition operator for amplitudes in which
the $D$ spectator quark appears in the final state and ${\cal T}_B$ is
the transition operator for annihilation amplitudes which are
discussed in subsection \ref{subsec:anni}. We write ${\cal T}_A$ as
\begin{eqnarray}\label{Toper}
   {\cal T}_A&=&  \sum_{p=d,s} \lambda_p\,
\left(a^P_1
    (\bar p c)_{V-A} \otimes (\bar u p)_{V-A} + a^P_2 
   (\bar u c)_{V-A} \otimes (\bar p  p)_{V-A} \right)\nonumber\\
   &&+      a^P_3  \sum_{\!q}\, (\bar u c)_{V-A} \otimes
    (\bar q q)_{V-A}  + a^P_4  \sum_{\!q}\, (\bar q c)_{V-A} \otimes
    (\bar u q)_{V-A}  \nonumber\\
    &&+ a^P_5  \sum_{\!q}\, (\bar u c)_{V-A} \otimes
    (\bar q q)_{V+A} +  a^P_6 (-2) \sum_{\!q}\, (\bar q c)_{S-P} \otimes
    (\bar u q)_{S+P} \,,
\end{eqnarray}
where $P=K,\,\pi$ for $D \to KK,\,\pi\pi$ decays, respectively, a
summation over $q=u,d,s$ is implied and $\lambda_p = V_{cp}^* V_{up}$.
Fierzing of $Q_5\,,Q_6$ gives rise to the $(S-P)(S+P) $ term.  The
second pair of quarks in each term produces a final state meson
($P_2$), and the outgoing quark in the first pair combines with the
spectator quark to form a final state meson ($P_1$).  The $\otimes$
indicates that the matrix element of the corresponding operator in
${\cal T}_A$ is to be evaluated in the factorized form:
\beqa
\langle P_1 P_2 |j_1\otimes j_2|D\rangle&\equiv&
\langle P_1 |j_1|\bar D\rangle\,\langle P_2 |j_2|0\rangle  \nonumber   \\
&=&\begin{cases} -i  c  {\cal A}_{P }
\,,&~ \protect{{\rm for~~}j_1\otimes j_2=(V-A) \otimes (V\mp A),}   \cr
-i c \,r_\chi \, {\cal A}_{P }\,,&
~ \protect{{\rm for~~}j_1\otimes
j_2= -2 ( S-P) \otimes (S+P).} \end{cases}
\label{factdef}
\eeqa
The $c$ coefficients are products of factors of $\pm 1$,
$\pm1/\sqrt{2}$, 
which depend on the flavor structures of the mesons, and
\begin{equation}
\label{Am1m2}
{\cal A}_{P} = i {G_F \over \sqrt{2}} \, 
(m_D^2 - m_{P }^2 ) F_0^{D \to P } (m_{P}^2 )\, f_{P } \,,
\end{equation}
where $F_0^{D \to P }$ is the $D \to P$ transition form factor and
$f_{P}$ is the decay constant.  The factor $r_\chi$ appearing in the
scalar matrix elements is given by
\beq
\label{rchi}
r_\chi=\frac{\mu_P}{m_c}, \qquad
\mu_P = \frac{2 m_K^2}{m_s + m_q }= \frac{2 m_\pi^2}{m_u + m_d},
\qquad m_q = \frac{m_u + m_d}{2}.
\eeq

The $a^P_i$ coefficients in general contain the contributions from
naive factorization, penguin contractions, vertex corrections, and
hard spectator interactions.  We only consider explicitly the naive
factorization contributions for $Q_{1,..,6}$, and the penguin
contraction for $Q_{8g}$ \cite{Beneke:2001ev}.  We therefore obtain
($N_c = 3$ and $P=K,\pi$)
\begin{eqnarray}
\label{ai}
a^P_1  & =& C_1 + \frac{C_2}{N_c},\qquad
a^P_2 = C_2 +\frac{C_1}{N_c}, \qquad
a^P_3 = C_3 + \frac{C_4}{N_c}, \qquad
a^P_5 = C_5 + \frac{C_6}{N_c} ,\nonumber \\
a^P_4 &=& C_4 + {C_3 \over N_c }  -  
 {C_F \alpha_s \over 2 \pi N_c }  
 C_{8g}  \int_0^1 {\phi_{P}(x) \over x}  dx,\qquad
a^P_6 =  C_6 + {C_5 \over N_c}  -  {C_F \alpha_s \over 2 \pi N_c } C_{8g}\,.  
\end{eqnarray}
where $C_F=(N_c^2 -1)/(2 N_c )$.
$\phi_P(x)$ is the leading-twist light-cone meson distribution
amplitude for meson $P$.  For simplicity we consider asymptotic
distribution amplitudes, in which case
\beq
a_i^K = a_i^\pi \equiv a_i, \qquad \int_0^1 {\phi_{P} (x) \over x }  dx=3.
\eeq
In that case the only sources of $SU(3)_F$ breaking are the
form-factors and decay constants. $A_{\rm NF}$, the naive factorization
amplitudes for $D \to KK/\pi\pi$ are then given by
\begin{eqnarray}
\label{leadingpoweramps}
A_{\rm NF}(D \to K^0 \overline{K}^0) &=&0, \nonumber \\
A_{\rm NF}(D^0 \to K^+ K^-  ) & =&A_{\rm NF}(D^+ \to K^+ K^0) =  
(\lambda_{s } \,a_1 + a_4 + r_\chi a_6   ) {\cal A}_{K  },\nonumber \\
A_{\rm NF}(D^0 \to \pi^+ \pi^-)&=&
 (\lambda_{d}\,a_1 + a_4 + r_\chi a_6) {\cal A}_{\pi},\nonumber\\
-\sqrt{2}\,A_{\rm NF}(D^+ \to \pi^+ \pi^0)&=&
   \lambda_{d}\,(a_1  + a_2){\cal A}_{\pi}\nonumber\\
A_{\rm NF}(D^0  \to  \pi^0 \pi^0 ) &=&
 (-\lambda_{d } \,a_2 + a_4 + r_\chi a_6   ) {\cal A}_{\pi  } \,.
\end{eqnarray}
The decay $D \to K^0 \overline{K}^0 $ only proceeds via annihilation
and thus vanishes in (\ref{leadingpoweramps}).
The standard model amplitudes are given in terms of $a_{1,2}$, and the
new physics amplitudes are given in terms of $a_{3,..,6}$. 

Note that there are no strong phase differences at this point between
the SM and NP amplitudes.  However, large power-corrections (or
final-state interactions) could generate them.  As we argue below, the
measured decay widths point to a large role for such
infrared dominated physics, especially annihilation.  Thus, the large
strong-phase differences that would be necessary to obtain $a_f^d \sim
r_f$ are well motivated.

In our numerical estimates we take $m_c (m_c ) =1.64 $ GeV and $m_s =
110$ MeV, $m_u + m_d = 9$ MeV at $\mu = 2$ GeV.  The scale at which
the Wilson coefficients and $r_\chi $ are evaluated is varied within
the range $\mu \approx 1-2$ GeV.  At $\mu \approx m_c $ we obtain
$r_\chi \approx 2.5$, $a_1 \approx 1.05$, and $a_2 \approx 0.05$ at
next-to-leading order. (For simplicity we ignore the $m_b$ quark mass
threshold, taking $n_f =5$ and $\Lambda_{\rm QCD} = 225$ MeV).  With
regards to the form factors, the BES Collaboration has measured
\cite{BESF0}
\begin{equation}
\label{FDP}
F_+ (0 )^{D \to K} = 0.78 \pm 0.04 \pm 0.03, \qquad
F_+ (0 )^{D \to \pi} =  0.73 \pm 0.14 \pm 0.06.
\end{equation}
(A recent lattice determination obtains $F_+ (0 )^{D \to K} = 0.73 \pm
0.03 \pm 0.07$ \cite{latticeFDK}.)  The values of $F^{D \to K,\pi}_0
(0 )$ entering ${\cal A}_{K,\pi}$ follow from the kinematical
constraint $F_0 (0) = F_+ (0)$.  Small shifts in $F_{0} (q^2) $ due to
$q^2 \lsim m_{K}^2$ are negligible. Our estimates are obtained by
varying the measured values of $F^{D\to K,\pi}_0 (0)$ in their $\pm
1\sigma$ ranges quoted in (\ref{FDP}). 

The decay rates are given by
\beq
\Gamma (D\to PP) = { |A(D \to PP)|^2  \over 16 \pi m_D } \, 
\sqrt{1-{4m^2_P\over m_D^2}}, \qquad A=A_{\rm NF}+A_{\rm ann}. 
\eeq
Taking $A_{\rm ann}=0$ we get the following naive factorization decay
widths within the SM,
\beqa
\label{KKBRsatLO}
\Gamma(D^0 \to K^+K^- ) &=&\Gamma(D^+ \to K^+{\overline K^0} )= (4.6 -
6.5 )\times 10^{-6}\ {\rm eV}\,,
\nonumber\\
\Gamma(D^0 \to \pi^+\pi^- ) &=& (2.6-6.5)\times 10^{-6}\ {\rm
eV}\,,\nonumber \\
\Gamma(D^0 \to \pi^0\pi^0 ) &=& (0.1   -  0.3)\times 10^{-6}\ {\rm
eV}\,, \nonumber \\
\Gamma(D^+ \to \pi^+\pi^0 ) &=& (1.3   - 3.5)\times 10^{-6}\ {\rm
eV}\,.
\eeqa

%%%%%%%%%%%%%%%%%%%%%%%%%%%%%%%%%%%%%%%%%%%%%
\subsection{Annihilation}
\label{subsec:anni}
Adapting \cite{Beneke:2003zv} to $D\to PP$ decays, the
annihilation matrix elements can be organized in terms of flavor
operators of the form $B([\bar q_{P_1} q_{P_1}] \,[\bar q_{P_2}
q_{P_2}]\,[\bar q_s c ] )$, where $q_s$ denotes the spectator
antiquark in the $D$-meson.  The matrix element of a B-operator is
defined as
\begin{equation}
\langle P_1 P_2 |\,B([..][..][..])| D  \rangle =  
c\, {\cal B}_{P} \,,~~~{\rm with}~~{\cal B}_{P} =  i{G_F \over \sqrt{2}}f_D f^2_{P}\, 
\end{equation}
whenever the quark flavors of the three brackets match the three
mesons, respectively. The notations are as in Eq. (\ref{factdef}).
The transition operator for the annihilation contributions of
$Q_{1,..,6}\,,Q_{8g}$ in Eq. (\ref{Top}) can be parametrized in full
generality as
\beqa
\label{TBops}
{\cal T}_B &=& \sum_{p=d,s} \lambda_p \, \left(\, \sum_{q^\prime} b^P_{1
q^\prime } \, B ([\bar p q^\prime ][\bar q^\prime p][\bar u c]) +
\delta_{pd}\, \sum_{q^\prime} b^P_{2 q^\prime} \, B( [\bar u q^\prime
][\bar q^\prime d][\bar d c])\, \right)\nonumber\\ &&+ \sum_{q,q^\prime} b^P
_{3 q^\prime} \,B([\bar u q^\prime]\bar q^\prime q][\bar q c])+
\sum_{q,q^\prime}b^P_{4q^\prime} B( [\bar q q^\prime]\bar q^\prime
q][\bar u c])\,,
\eeqa
where $q,q^\prime = u,d,s$. Here $q^\prime$ denotes the flavor of the
``popped'' quark-antiquark pair from gluon splitting, $g \to \bar
q^\prime q^\prime $.  Isospin symmetry implies $b^\pi_{iu}=b^\pi_{id} \equiv
b^\pi_i $, $b_{iu}^{K} = b_{id}^{K}$, and $U$-spin symmetry would
further imply $b_{is}^{K} = b_{id}^{K}= b^\pi_i $.  $b^P_{1,2} $
receive contributions from the SM current-current operators, and
$b^P_{3,4}$ from NP via the QCD penguin and dipole operators.

Using isospin all of the $b_i$ coefficients can be expressed in terms
of $D \to P^+ P^- , K^+ \overline {K}^0 $ effective operator
annihilation matrix elements. For the SM operators we have
\beqa
\label{biQi}
{\cal B}_P \,b^P_{1q^\prime } &= & C_1 \,\langle P^+ P^- | (\bar p_\alpha p_\beta )_{V-A} \otimes^A (\bar u_\beta c_\alpha)_{V-A}| D^0 \rangle
+  \nonumber\\&& 
C_2 \, \langle P^+ P^- | (\bar p  p)_{V-A} \otimes^A (\bar u c)_{V-A}|D^0 \rangle,\nonumber\\
{\cal B}_K b^K_{2s} &=& C_1\, \langle K^+ \overline{K}^0 |(\bar u d )_{V-A} \otimes^A (\bar d c)_{V-A}|D^+ \rangle
+  \nonumber\\&&
C_2\,  \langle K^+ \overline{K}^0 | (\bar u_\alpha d_\beta )_{V-A}
\otimes^A (\bar d_\beta c_\alpha)_{V-A}|D^+ \rangle.
\eeqa
For the NP operators we have
\beqa
{\cal B}_P b^P_{3 q^\prime }&=& C_3  \, \langle P^+ P^- | (\bar
u_\alpha u_\beta  )_{V-A}  \otimes^A (\bar u_\beta c_\alpha
)_{V-A}|D^0  \rangle +  \nonumber\\&&
C_4 \,\langle P^+ P^- | (\bar u u  )_{V-A}  \otimes^A (\bar u c)_{V-A}
|D^0 \rangle + \nonumber\\&& 
C_5  \,\langle P^+ P^- | \!-2\, (\bar u_\alpha  u_\beta )_{S+P} \otimes^A (\bar u_\beta c_\alpha  )_{S-P}|D^0 \rangle
+  \nonumber\\&&
C_6 \,  \langle P^+ P^- | \!-2\, (\bar u u )_{S+P} \otimes^A (\bar u c )_{S-P} |D^0  \rangle, \nonumber\\
{\cal B}_P b^P_{4 q^\prime} &= & C_3 \, \langle P^+ P^- | (\bar q q  )_{V-A}  \otimes^A (\bar u c )_{V-A}|D^0 \rangle + \nonumber\\&&
C_4\, \langle P^+ P^- |(\bar q_\alpha q_\beta  )_{V-A}  \otimes^A (\bar u_\beta c_\alpha )_{V-A}| D^0 \rangle \nonumber \\
&+  & C_5\, \langle P^+ P^- |  (\bar q q  )_{V+A}  \otimes^A (\bar u c )_{V-A}|D^0 \rangle+ \nonumber\\&&
C_6\, \langle P^+ P^- | (\bar q_\alpha q_\beta  )_{V+A}  \otimes^A (\bar u_\beta c_\alpha )_{V-A}|D^0 \rangle \,.
\eeqa
The annihilation product $j_1 \otimes^A j_2 $ means that $j_2 $
destroys the $D$ meson, and $j_1$ creates a quark and an antiquark
which end up in different mesons.  The choices of $p $ and $q $ among
$(d,s)$ and $(u,d,s)$, respectively, are fixed by the values taken by
$P$ and $q^\prime$.  In $b^P_{4 q^\prime}$ the
$\langle {j_1}_{V-A} \otimes^A {j_2}_{V-A} \rangle $ and $\langle
{j_1}_{V+A} \otimes^A {j_2}_{V-A} \rangle $ matrix elements are equal
because parity implies $\langle P^+ P^- | (\bar q q )_{V-A} | g_1
...g_n \rangle =\langle P^+ P^- | (\bar q q )_{V+A} | g_1 ...g_n
\rangle $.  Finally, we point out that $Q_{8g}$ also contributes to
$b^P_{3q^\prime}$ and $b^P_{4q^\prime}$.  A discussion of the
theoretical uncertainty for dipole operator amplitudes due to the
annihilation topology is left for future work.

Assuming isospin, the annihilation amplitudes are given by 
\begin{eqnarray}
\label{Annamps}
{A}_{\rm ann} (D \to \pi^+ \pi^- )&=&
 {\cal B}_{\pi}\, (\lambda_d b^\pi_1 + b^\pi_3 + 2\,b^\pi_4 ) \nonumber\\ 
{A}_{\rm ann} (D\to K^+ K^- ) &=& 
{\cal B}_{K}\, (\lambda_s b^{K}_{1u}+ b^{K}_{3s} + b^{K}_{4s} + b^{K}_{4u}) \nonumber\\ 
{A}_{\rm ann} (D \to \pi^0 \pi^0 ) & = & 
{A}_{\rm ann} (D \to \pi^+ \pi^- )\nonumber \\ 
{ A}_{\rm ann} (D \to K^0 \overline{K}^0 ) & = & 
{\cal B}_{K}\, (\lambda_s[b^{K}_{1d}-b_{1s}^{K}]+b_{4d}^{K}+b_{4s}^{K})\nonumber\\
{A}_{\rm ann} (D \to \pi^+ \pi^0 )&=& 0 \nonumber \\ 
{A}_{\rm ann}(D \to K^+ \overline{K}^0)&=& 
{\cal B}_{K}\,( \lambda_s b_{2s}^{K} + b_{3s}^{K} )\,.
\end{eqnarray}
Note that in the $U$-spin limit $A_{\rm ann}(K^+ K^-)=A_{\rm
ann}(\pi^+ \pi^-)$ and, neglecting the penguin operators, $A_{\rm ann}
(K^0 \overline{K}^0) =0$.

In order to estimate the $b_i$'s we make use of the tree-level one
gluon exchange approximation \cite{Beneke:2001ev,Beneke:2003zv}. 
In general, factorizable contributions to $\langle j_1 \otimes^A j_2
\rangle$, of the form $\langle P_1 P_2 | j_1 |0 \rangle \langle
0| j_2 |D\rangle$, vanish for the $(V\pm A)\otimes^A (V-A)$ matrix
elements by the equations of motion. Therefore, many of the matrix
elements in Eq. (\ref{biQi}) vanish in the one-gluon approximation. We
further simplify our discussion by taking asymptotic meson light-cone
distribution amplitudes. Then, the number of independent building blocks
appearing in Eq. (\ref{biQi}) reduces to two
\cite{Beneke:2001ev,Beneke:2003zv},
\beqa
\label{Ai}
A_1^i &=& \langle P^+ P^- | (\bar q_{\alpha} q_{\beta} )_{V\mp A} \otimes^A (\bar u_\beta  c_\alpha)_{V-A}| D  \rangle /  {\cal B}_P   =  \frac{C_F}{N^2 }    \pi \alpha_s \left[18 \left(X-4 + \frac{\pi^3}{3} \right) + 2 r_\chi^2 X^2 \right]\nonumber\\
A_3^f &=&  \langle P^+ P^- | \!-2\, (\bar q q )_{S+P} \otimes^A (\bar u  c )_{S-P} |D^0  \rangle/{\cal B}_P = \frac{C_F}{N }  12 \pi \alpha_s r_\chi (2 X^2 - X )\,.
\eeqa
The superscripts $i \,(f)$ denote a gluon exchanged from the initial
(final-state) quarks in the four-quark operator.  $X$ represents
an incalculable infrared logarithmically divergent quantity which
signals a breakdown in short/long distance factorization.  It is a necessary
model-dependent input in the one-gluon approximation.  
For simplicity, we take $X$ to be universal. Adopting the model of
\cite{Beneke:2001ev}, $X$ is parametrized as
\begin{equation}
\label{Xrhophi}
X = \log(m_D /\Lambda_h ) (1+ \rho e^{i \phi }).
\end{equation}
$\Lambda_h \sim 500$ MeV is a hadronic mass scale corresponding to
some physical infrared cutoff, $\phi$ allows for the presence of an
arbitrary strong phase from soft rescattering, and $\rho$ parametrizes
our ignorance of the magnitudes of these amplitudes.  With our
assumptions, we get
\beqa
\label{Aibi}
b^P_{1q^\prime} &= & C_1 A_1^i\,,\qquad 
b^P_{3 q^\prime}=C_3 A_1^i+\left(C_6+{C_5\over
N_c}\right)A_3^f\nonumber\\ 
b^K_{2s} &=& C_2 A_1^i\,,\qquad
b^P_{4 q^\prime}=(C_4 + C_6 )A_1^i \,.
\eeqa
The strong color-suppression $b^K_{2s}\ll b^K_{1q^\prime }$ may be an 
artifact of the one-gluon approximation, as beyond it the contribution of the
matrix-element of $Q_1$ to $b^K_{2s}$ does not vanish.

In our numerical evaluation we use $\alpha_s$ and $r_\chi$ in
Eq. (\ref{Ai}) at a scale $\mu_h \approx 0.7 $ GeV, corresponding to
$\alpha_s \approx 1$ (reflecting the infrared dominance of these
matrix elements).  The Wilson coefficients are evaluated at a scale
$\mu = m_c (m_c)$.  For $f_D$ we take the central
value of the CLEO-c measurement, $f_D = 223 \pm 17 \pm 3$ MeV
\cite{CLEOcfD}.

%%%%%%%%%%%%%%%%%%%%%%%%%%%%%%%%%%%%%%%%
\subsection{Comparison with data}
In order to estimate the value of the model parameters we compare the
prediction with the measured widths \cite{PDG}
\beqa
\label{KKBRs}
\Gamma(D^0 \to K^+K^- ) &=& (6.16 \pm 0.16)\times 10^{-6}\ {\rm eV}\,,
\nonumber\\
\Gamma(D^0 \to \pi^+\pi^- ) &=& (2.19 \pm 0.05)\times 10^{-6}\ {\rm
eV}\,,\nonumber \\
\Gamma(D^0 \to K^0\overline{K}^0) &=& 
   (1.19 \pm 0.22)\times 10^{-6}\ {\rm eV}\,,\nonumber\\
\Gamma(D^0 \to \pi^0\pi^0 ) &=& (1.27 \pm 0.13)\times 10^{-6}\ {\rm
eV}\,, \nonumber \\
\Gamma(D^+ \to K^+\overline{K}^0) &=& 
   (3.75 \pm 0.24)\times 10^{-6}\ {\rm eV}\,,\nonumber\\
\Gamma(D^+ \to \pi^+\pi^0 ) &=& (0.81 \pm 0.04)\times 10^{-6}\ {\rm
eV}\,.
\eeqa
We always assume that the NP amplitudes are small, so the
above measured rates are given by the SM.

To leading order in $1/m_c$, only spectator diagrams contribute to the
various decay amplitudes. Comparing the naive factorization
predictions, Eq. (\ref{KKBRsatLO}), with the measured values we see that
they are in disagreement. In particular, $\Gamma (D^0\to{K^0
\overline{K}^0})$ and $\Gamma(D^0\to \pi^0 \pi^0)$ are considerably larger than the
naive factorization predictions. 
The predicted rates for the $K^+ K^- $, $K^+ \overline{K}^0$, $\pi^+
\pi^- $, and $\pi^+ \pi^0$ modes are of the correct order of magnitude.
However, rather than being equal as expected at leading power, $\Gamma
(K^+ K^- )$ is approximately twice $\Gamma (K^+ K^0 )$ .

The disagreement points to a substantial role for annihilation.
The magnitudes of the observed amplitudes imply that 
\beq
|A(K^0 \overline{K}^0)| \sim \frac12 |A(K^+ K^-)|.
\eeq
$A(K^+ K^-)$ has contributions from both naive factorization and
annihilation amplitudes. $A(K^0 \overline{K}^0)$ on the other hand is
pure annihilation and it vanishes in the SM in the U-spin limit. We
therefore expect that
\beq
|A_{\rm ann}(K^0 \overline{K}^0)| \lsim |A_{\rm ann}(K^+ K^-)|.
\eeq
Since naive factorization predicts the right orders of magnitude for the $P^+
P^- $ widths, we expect that the annihilation and naive factorization amplitudes are of
same order for $K^+ K^-$.  The same should be true for $\pi^+ \pi^-$ based on any reasonable pattern
for $SU(3)_F$ breaking.
We therefore write schematically
\begin{equation}
\label{AKTK}
\frac{A^{\rm SM}_{\rm ann}}{A^{\rm SM}_{\rm NF}}  \sim {\langle P^+
P^- | (\bar p_{\alpha} p_{\beta} )_{V- A} \otimes^A (\bar u_\beta
c_\alpha)_{V-A}| D^0  \rangle   \over \langle P^+ P^- | (\bar p c
)_{V- A} \otimes (\bar u p)_{V-A}| D^0  \rangle } \sim 1.
\end{equation}

Next we try to estimate the size of the NP annihilation amplitudes. We
use the one-gluon exchange model discussed above. We see that 
Eq. (\ref{AKTK}) is reproduced with $X\sim 5$ in
Eq. (\ref{Ai}).\footnote{$X\approx 5$ arises, e.g.,  for $\rho \sim 3$ and
$\phi \sim 0$ in Eq. (\ref{Xrhophi}). It is worth mentioning that
similar values of $\rho$ are required in order to account for the $e^+
e^- \to P^+ P^-$ cross sections at $\sqrt{s} \approx 3.7$ GeV
\cite{dk}.}  Using $X\sim 5$ in Eq. (\ref{Ai}) for $A_3^f$ we can
estimate the size of the NP annihilation amplitude. We find that the
chirally-enhanced QCD penguin annihilation amplitude is much larger
than the corresponding spectator amplitude. They also tend to dominate
the total penguin annihilation and total penguin spectator amplitudes, respectively, in NP models.
Schematically, we write this as
\beq
\label{AKTK-NP}
\frac{A^{\rm NP}_{\rm ann}}{A^{\rm NP}_{\rm NF}} \sim   { \langle P^+ P^- |  (\bar u u )_{S+P} \otimes^A (\bar u c )_{S-P} |D^0  \rangle \over    \langle P^+ P^- |  (\bar q c )_{S-P} \otimes (\bar u q )_{S-P} |D^0  \rangle }
\sim 5.
\eeq
The large ratio implies that new QCD penguin amplitudes in $D\to PP$
decays could receive an order-of-magnitude enhancement from
annihilation. This is demonstrated in the numerical example of
Fig.~\ref{fig:deltaLLfig}(c), where the annihilation matrix elements
are included as above with $X\approx 5$ ($\rho=3\,,\phi = 0$).

Given the crude nature of the one-gluon exchange approximation this
should only be taken as an indication of the theoretical uncertainty
due to QCD penguin operator annihilation.  
A similar analysis of the
theoretical uncertainty for the dipole operator matrix element due to
the annihilation topology is left for future work.

%%%%%%%%%%%%%%%%%%%%%%%
%%%%%%%%%%%%%%%%%%%%%%%%%%%%%%%%%%%%%%%%%%%%%%%%%
\section{QCD penguin and dipole operators in SUSY}

We study contributions to the QCD penguin and dipole operator Wilson
coefficients arising from up squark-gluino loops.  For simplicity, we
work in the squark mass-insertion approximation where to first
approximation the squark masses are degenerate with mass $\tilde{m}$.
In particular, we consider the contributions of the up-squark mass
insertions $\delta_{LL}$ and $\delta_{LR}$ to $C_{3,..,6}$, $C_{8g}$.
(Since in our case $\delta_{LR} \ll 1$ and $\delta_{LL}\lsim 1$, the
mass insertion approximation works very well for $\delta_{LR}$ and
only provides rough estimates for $\delta_{LL}$.)  The expressions for
the SUSY Wilson coefficients are given at the scale $\mu \sim m_{SUSY}
$ by~\cite{ko01}
\begin{eqnarray}
\label{susyCidecay}
   C_3 & = & 
   -\frac{\alpha_s^2}{ 2 \sqrt{2} G_F \tilde{m}^2 }
   \left(-\frac{1}{9} B_1(x)-\frac{5}{9} B_2(x)-\frac{1}{18} P_1(x)
   -\frac{1}{2} P_2(x) \right) 
\delta_{LL}  \nonumber \\
   C_4 & = & 
   -\frac{\alpha_s^2}{  2 \sqrt{2} G_F \tilde{m}^2}
   \left(-\frac{7}{3} B_1(x)+\frac{1}{3} B_2(x)+\frac{1}{6} P_1(x)
   +\frac{3}{2} P_2(x) \right) 
   \delta_{LL}  \nonumber \\
   C_5  & = & 
   -\frac{\alpha_s^2}{ 2 \sqrt{2} G_F \tilde{m}^2 }
   \left(\frac{10}{9} B_1(x)+\frac{1}{18} B_2(x)-\frac{1}{18} P_1(x)
   -\frac{1}{2} P_2(x) \right) 
  \delta_{LL}  \nonumber \\
   C_6 & = & 
   -\frac{\alpha_s^2}{ 2 \sqrt{2} G_F \tilde{m}^2}
   \left(-\frac{2}{3} B_1(x)+\frac{7}{6} B_2(x)+\frac{1}{6} P_1(x)
   +\frac{3}{2} P_2(x) \right) \delta_{LL}
    \nonumber \\
   C_{8 g}& = & - {2 \pi \alpha_s \over 
   \sqrt{2} G_F \tilde{m}^2 }
   \left[ \delta_{LL} \left( {3\over 2} M_3 (x) - 
   {1\over 6} M_4 (x) \right) \right.
   \nonumber  \\
   & & \left.
   +  \delta_{LR} \left( {m_{\tilde{g}} \over m_c} \right) 
   ~{1\over 6}~\left( 4 B_1 (x) - 9 x^{-1} B_2 (x) \right) \right], 
\end{eqnarray}
where $x \equiv ( m_{\tilde{g}} / \tilde{m} )^2$, and the loop
functions can be found in Ref.~\cite{ko01}.  (The mass insertions
$\delta_{RR}$ and $\delta_{RL}$ generate the opposite chirality
operators $\tilde Q_i$).  For simplicity, we evaluate the above
Wilson coefficients at $\mu = m_t $, and evolve them to $\mu = m_c $
at LO.

The $\Delta C=2$ effective Hamiltonian, $H_{\rm eff}^{\Delta C=2 }$, for supersymmetric up squark-gluino box graph contributions to 
$D-\overline D $ mixing 
is given in Eqs. (\ref{eq:eh}), with
\begin{eqnarray}
O_1  &=&   \bar{u}_L^{\alpha} \gamma_\mu c_L^{\alpha}~
           \bar{u}_L^{\beta}  \gamma^\mu c_L^{\beta}\,, \qquad
O_2 =   \bar{u}_R^{\alpha} c_L^{\alpha}~\bar{u}_R^{\beta} c_L^{\beta}\,, 
\qquad
O_3  =  \bar{u}_R^{\alpha} c_L^{\beta} ~\bar{u}_R^{\beta} c_L^{\alpha}\,, 
\nonumber  \\
O_4 & = &  \bar{u}_R^{\alpha} c_L^{\alpha}~\bar{u}_L^{\beta} c_R^{\beta}\,, 
\qquad
O_5  =   \bar{u}_R^{\alpha} c_L^{\beta} ~\bar{u}_L^{\beta} c_R^{\alpha}\,.
\label{eq:db2ops}
\end{eqnarray}
The $D -\overline{D}$ mixing amplitude is given by
$M_{12}^D = \langle D | H_{\rm eff}^{\Delta C = 2} |\overline{D} \rangle / 2 m_D $, where 
$\Delta m_D = 2 M_{12}^D = x \Gamma_D $.
In the squark mass insertion approximation 
the SUSY Wilson coefficients for the operators $O_i$ are given by \cite{antichi},
\begin{eqnarray}
c_1 & = & 
- {\alpha_s^2 \over 216 \tilde{m}^2}~\left( 24 x f_6 (x) + 66 
\tilde{f}_6 (x)\right) ~\left( \delta_{13}^d \right)_{LL}^2 \,
\nonumber  \\
c_2 & = & - {\alpha_s^2 \over 216 \tilde{m}^2}~204 x f_6 (x) 
~ \delta_{RL}^2 \,
\nonumber  \\
c_3 & = & {\alpha_s^2 \over 216 \tilde{m}^2}~36 x f_6 (x) 
~\delta_{RL}^2 \,
\nonumber  \\
c_4 & = & - {\alpha_s^2 \over 216 \tilde{m}^2}~\left[~ 
\left( 504 x f_6 (x) - 72 \tilde{f}_6 (x) \right)~
\delta_{LL} \delta_{RR}
  - 132 \tilde{f}_6 (x) ~
\delta_{LR}\delta_{RL} ~\right]\,
\nonumber  \\
c_5 & = & - {\alpha_s^2 \over 216 \tilde{m}^2}~\left[~ 
\left( 24 x f_6 (x) + 120 \tilde{f}_6 (x) \right)~
\delta_{LL}  \delta_{RR} - 180 \tilde{f}_6 (x) ~
 \delta_{LR}  \delta_{RL} ~\right] .
\label{dD2wilsonSUSY}
\end{eqnarray}
The other Wilson coefficients $\tilde{c}_{i=1,2,3}$ are 
obtained from $c_{i=1,2,3} $ by exchange of $L\leftrightarrow R$.
The loop functions
are given by
\begin{eqnarray}
f_6 (x) & = & {6 ( 1 + 3 x ) \ln x + x^3 - 9 x^2 - 9 x + 17 \over 
               6 ( x - 1 )^5 } ,
\nonumber \\
\tilde{f}_6 (x) & = & {6 x ( 1+x)\ln x - x^3 - 9 x^2 + 9 x + 1 \over
               3 ( x - 1 )^5 } .
\end{eqnarray}
Again, the Wilson coefficients are evaluated at $\mu = m_t $ and evolved down to $\mu \approx m_c $ at LO  \cite{bagger}.
%As in decays we limit our discussion to effects of the $\delta_{LL}$ 
%and $\delta_{LR}$ insertions, which only yield contributions to $O_{1,2,3}$.
For simplicity, we use the vacuum insertion approximation for the operator matrix elements, 
\begin{eqnarray}
\langle D | O_1  | \overline{D} \rangle & = &
{2\over 3}~m_{D}^2 f_{D}^2 ,
\nonumber  \\
\langle D | O_2  | \overline{D} \rangle & = &
-{5\over 12}~\left( { m_{D} \over m_c  + m_u  } \right)^2~
m_{D}^2 f_{D}^2 ,
\nonumber  \\
\langle D | O_3  | \overline{D} \rangle & = &
{1\over 12}~\left( { m_{D} \over m_c  + m_u  } \right)^2~
m_{D}^2 f_{D}^2 ,\nonumber\\
\langle D | O_4  | \overline{D} \rangle & = &
{1\over 2}~\left( { m_{D} \over m_c  + m_u  } \right)^2~
m_{D}^2 f_{D}^2,\nonumber\\
\langle D | O_3  | \overline{D} \rangle & = &
{1\over 6}~\left( { m_{D} \over m_c  + m_u  } \right)^2~
m_{D}^2 f_{D}^2 \,.
\label{eq:dD2me}
\end{eqnarray}

%%%%%%%%%%%%%%%%%%%%%%%%%%%%%

\end{document}